\documentclass{emulateapj}
\usepackage{hyperref,xcolor}

\slugcomment{}

\shorttitle{Simulated spectra of high-redshift galaxies}
\shortauthors{Zackrisson et al.}

\begin{document}

\title{The Spectral Evolution of the First Galaxies. III. Simulated James Webb Space Telescope Spectra of Reionization-Epoch Galaxies with Lyman Continuum Leakage}

\author{Erik Zackrisson\altaffilmark{1}$^*$, Christian Binggeli\altaffilmark{1}, Kristian Finlator\altaffilmark{2}, Nickolay Y. Gnedin\altaffilmark{3,4,5},\\ Jan-Pieter Paardekooper\altaffilmark{6}, Ikkoh Shimizu\altaffilmark{7}, Akio K. Inoue\altaffilmark{8}, Hannes Jensen\altaffilmark{1}, Genoveva Micheva\altaffilmark{9},\\ Sadegh Khochfar\altaffilmark{10} \& Claudio Dalla Vecchia\altaffilmark{11,12}}
\altaffiltext{*}{E-mail: erik.zackrisson@physics.uu.se}
\altaffiltext{1}{Department of Physics and Astronomy, Uppsala University, Box 515, SE-751 20 Uppsala, Sweden}
\altaffiltext{2}{New Mexico State University, MSC 4500, Las Cruces, NM 88003, USA}
\altaffiltext{3}{Particle Astrophysics Center, Fermi National Accelerator Laboratory, Batavia, IL 60510, USA}
\altaffiltext{4}{Department of Astronomy \& Astrophysics, The University of Chicago, Chicago, IL 60637 USA}
\altaffiltext{5}{Kavli Institute for Cosmological Physics and Enrico Fermi Institute, The University of Chicago, Chicago, IL 60637 USA}
\altaffiltext{6}{Universit\"at Heidelberg, Zentrum f\"ur Astronomie, Institut f¨ur Theoretische Astrophysik, Albert-Ueberle-Str. 2, 69120 Heidelberg, Germany}
\altaffiltext{7}{Department of Earth and Space Science, Osaka University, 1-1 Machikaneyama, Toyonaka, Osaka 560-0043, Japan}
\altaffiltext{8}{College of General Education, Osaka Sangyo University, 3-1-1, Nakagaito, Daito, Osaka 574-8530, Japan}
\altaffiltext{9}{University of Michigan, Department of Astronomy, 311 West Hall, 1085 S. University Ave, Ann Arbor, MI  48109-1107 USA}
\altaffiltext{10}{Institute for Astronomy, University of Edinburgh, Royal Observatory, Edinburgh, EH9 3HJ, UK}
\altaffiltext{11}{Instituto de Astrof\'isica de Canarias, C/ V\'ia L\'actea s/n, E-38205 La Laguna, Tenerife, Spain}
\altaffiltext{12}{Departamento de Astrof\'isica, Universidad de La Laguna, Av. del Astrof\'isico Francisco S\'anchez s/n, E-38206 La Laguna, Tenerife, Spain}

\begin{abstract}
Using four different suites of cosmological simulations, we generate synthetic spectra for galaxies with different Lyman continuum escape fractions ($f_\mathrm{esc}$) at redshifts $z\approx 7$--9, in the rest-frame wavelength range relevant for the James Webb Space Telescope (JWST) NIRSpec instrument. By investigating the effects of realistic star formation histories and metallicity distributions on the EW(H$\beta$)-$\beta$ diagram (previously proposed as a tool for identifying galaxies with very high $f_\mathrm{esc}$), we find that neither of these effects are likely to jeopardize the identification of galaxies with extreme Lyman continuum leakage. Based on our models, we expect that essentially all $z\approx 7$--9 galaxies that exhibit rest-frame EW(H$\beta$)$\lesssim 30$ \AA{} to have $f_\mathrm{esc}> 0.5$. Incorrect assumptions concerning the ionizing fluxes of stellar populations or the dust properties of $z>6$ galaxies can in principle bias the selection, but substantial model deficiencies of this type should at the same time be evident from offsets in the observed distribution of $z>6$ galaxies in the EW(H$\beta$)-$\beta$ diagram compared to the simulated one. Such offsets would thereby allow JWST/NIRSpec measurements of these observables to serve as input for further model refinement.  
\end{abstract}





\section{Introduction}
\label{intro}
Scenarios in which star-forming galaxies provide the photons that reionize the Universe at $z>6$ hinge on the assumption that hydrogen-ionizing (Lyman continuum, hereafter LyC) photons can readily escape from these objects and into the intergalactic medium. Current constraints place the required, time- and luminosity-averaged LyC escape fraction from galaxies in the reionization epoch at $f_\mathrm{esc}\approx 0.01$--0.2 \citep[e.g.][]{Hartley16,Mitra16,Sun16,Bouwens16}, but allowing for $f_\mathrm{esc}$ evolution with mass or redshift allows for considerably higher average values for certain subsets of reionization-epoch galaxies \citep[e.g.][]{Ferrara & Loeb,Price16}. Simulations also predict large galaxy-to-galaxy variations in $f_\mathrm{esc}$, with a few objects propelled to extreme levels of LyC leakage \citep[e.g.][]{Paardekooper13,Yajima et al.,Kimm & Cen,Paardekooper15,Cen & Kimm,Xu16}.

In the local Universe, all current direct detections of escaping LyC flux indicate $f_\mathrm{esc}\lesssim 0.1$ \citep{Bergvall06,Leitet et al.,Borthakur et al.,Izotov16a,Leitherer16,Izotov16b}, but a few objects with very high LyC escape fractions ($f_\mathrm{esc}>0.4$) have been reported at $z\approx 2$--3 \citep{Vanzella16,Matthee16,Shapley16}. In general, the physical mechanisms that provide the conditions for LyC photons to escape are not well constrained. In the language of simple gas/dust geometries, LyC escape can occur either through ionization-bounded nebulae with holes (``the picket-fence'' model) or through density-bounded nebulae \citep[for detailed discussions, see][]{Zackrisson13,Duncan15,Reddy16}. While some LyC-leaking galaxies do display the spectral signatures of density-bounded nebulae \citep{deBarros16}, \citet{Reddy16} argue that this scenario cannot apply to all cases.

Of course, what matters for cosmic reionization is the LyC escape fraction at $z\gtrsim 6$, where the opacity of the neutral IGM prevents the LyC to be measured directly \citep[e.g.][]{Inoue14}. Because of this, a number of observational methods aiming to provide indirect estimates of $f_\mathrm{esc}$ in the reionization epoch have recently been proposed \citep{Fernandez et al.,Jones et al.,Zackrisson13,Reddy16,Leethochawalit16}. 

Here, we continue the development of the method outlined by \citet{Zackrisson13} to measure LyC escape fractions at $z>6$. This technique exploits the fact that $f_\mathrm{esc}$ regulates the relative impact of nebular emission on the rest-frame ultraviolet/optical spectra of high-redshift galaxies \citep[e.g.][]{Zackrisson08}. Using spectral features that will be observable for galaxies at $z\approx 6$--9 using the NIRSpec instrument on the upcoming James Webb Space Telescope (JWST), \citet{Zackrisson13} argued that it should be possible to single out individual objects with very high escape fractions ($f_\mathrm{esc}\gtrsim 0.5$). Unlike other methods, this technique has modest signal-to-noise and spectral resolution requirements, and can therefore be applied to large numbers of high-redshift galaxies in upcoming JWST/NIRSpec surveys.

As the work of \citet{Zackrisson13} was based on a number of simplified assumptions concerning the properties of $z\gtrsim 6$ galaxies, we here replace the toy models used in that study with more realistic galaxies drawn from cosmological simulations. By generating mock galaxy spectra in the rest-frame ultraviolet/optical wavelength range relevant for JWST/NIRSpec, we show that neither the star formation histories nor the internal metal distributions of the simulated galaxies provide any insurmountable difficulties for the technique to identify galaxies with $f_\mathrm{esc}\gtrsim 0.5$. In a separate paper, \citep{Jensen et al.} use the same mock spectra to demonstrate that it may also in principle be possible to constrain the {\it typical} escape fraction fraction within samples of $\gtrsim 50$ galaxies studies with JWST/NIRSpec, even if the escape fraction is not very extreme ($f_\mathrm{esc}\lesssim 0.2$). The simulated spectral energy distributions (SEDs) for $z=7$--9 galaxies with different $f_\mathrm{esc}$, produced as part of the LYCAN (LYman Continuum ANalysis) project, are also made publicly available\footnote{LYCAN SEDs are publicly available at: \url{http://www.astro.uu.se/~ez/lycan/lycan.html}}.

The computational machinery and the associated assumptions used when generating the LYCAN spectra are described in Section~\ref{spectra}. In Section~\ref{diagnostic}, we discuss the impact of the star formation histories and metallicity distributions of simulated $z\approx 7$--9 galaxies on the diagnostic diagram of H$\beta$ emission-line equivalent width EW(H$\beta$) vs. UV slope $\beta$ introduced by \citet{Zackrisson13} to identify galaxies with extreme LyC escape fractions. 

While a realistic application of the proposed method to find such objects would likely involve all emission lines detectable in the rest-frame ultraviolet/optical \citep[as in][]{Jensen et al.}, the merits of focusing on this particular line is that this sidesteps many of the complexities related to the physical conditions of the gas. While several other emission lines are expected to be stronger than H$\beta$, hydrogen Balmer recombination lines\footnote{Unfortunately, H$\alpha$ redshifts out of NIRSpec range at $z>6.6$ -- otherwise this would likely have been the best choice} are comparatively insensitive to gas density, metallicity and ionization parameter.  Even in the case of H$\beta$, there are of course potential complications, which we discuss in Section~\ref{discussion}. Section~\ref{summary} summarizes our findings. 

\section{From cosmological simulations to synthetic galaxy spectra}
\label{spectra}
\subsection{Cosmological simulations of galaxy formation}
\label{simulations}
In this work, four independent suites of cosmological simulations are used to capture the star formation histories and metallicity distributions of galaxies in the reionization epoch. These are described in \citet[][hereafter F13]{Finlator13}, \citet[][hereafter S14]{Shimizu14}, \citet[][{\it Cosmic Reionization On Computers}; hereafter CROC]{Gnedin14} and \citet[][the {\it First Billion Years} simulations; hereafter FiBY]{Paardekooper13,Paardekooper15}. 

The F13, S14 and CROC simulations provide us with large numbers of synthetic galaxies with individual variations in star formation history and metallicity distributions, plus the opportunity to check the simulation-dependence of our final SEDs. However, neither of these simulation suites directly predict the LyC escape fraction $f_\mathrm{esc}$ of individual galaxies. This is instead provided by the FiBY simulations, albeit for a much smaller number of objects in the mass range relevant for JWST observations.

While these simulations provide output at a wide range of redshifts, we will in this paper focus on galaxies selected from simulation boxes at $z\approx 7$, except in the case of FiBY simulations, for which we have selected the most massive galaxies from the $z=6$, 7 and 8 boxes to increase the number of simulated galaxies with self-consistent $f_\mathrm{esc}$ predictions. Since the predicted evolution in galaxy properties (in terms of star formation history and internal metallicity distribution) is undramatic throughout this redshift range (see section~\ref{diagnostic_nodust}), no significant biases are expected to arise from this approach.

Throughout most of our analysis, we also impose lower limits on the total stellar mass of the simulated galaxies for which model SEDs are generated. This mass threshold serves the purpose of excluding galaxies too faint for reliable JWST/NIRSpec measurements of EW(H$\beta$) and $\beta$, and also to avoid artifacts related to the mass resolution of the simulations. \citet{Zackrisson13} estimated that these JWST/NIRSpec measurements require galaxies with total stellar masses $M_\mathrm{stars}\gtrsim 10^8$--$10^9 \ M_\odot$ in the absence of lensing and $M_\mathrm{stars}\gtrsim 10^7\ M_\odot$ in the case of strong lensing. Significant  resolution issues are also expected when the total stellar mass of the simulated galaxies approaches the stellar mass resolution element (the mass of the``star particles''; see section~\ref{SED_modelling}), as this makes the retrieved star formation history too uncertain for SED modelling. In the light of this, we only consider galaxies with $M_\mathrm{stars}\geq 10^7\ M_\odot$ for F13, CROC and FiBY and $M_\mathrm{stars}\geq 5\times 10^8\ M_\odot$ for S14. Because of differences in resolution and in the cosmological volumes used by the different simulation suites, these mass limits result in large variations in the number of objects for which LYCAN spectra are generated. At $z=7$, our criteria yield 874 galaxies for CROC, 406 for S14 and 106 for F13. In the case of FiBY, we get no more than 16 galaxies at $M_\mathrm{stars}\geq 10^7\ M_\odot$, even when combining data from $z=6$, 7 and 8.

\begin{figure*}
\centering
\plottwo{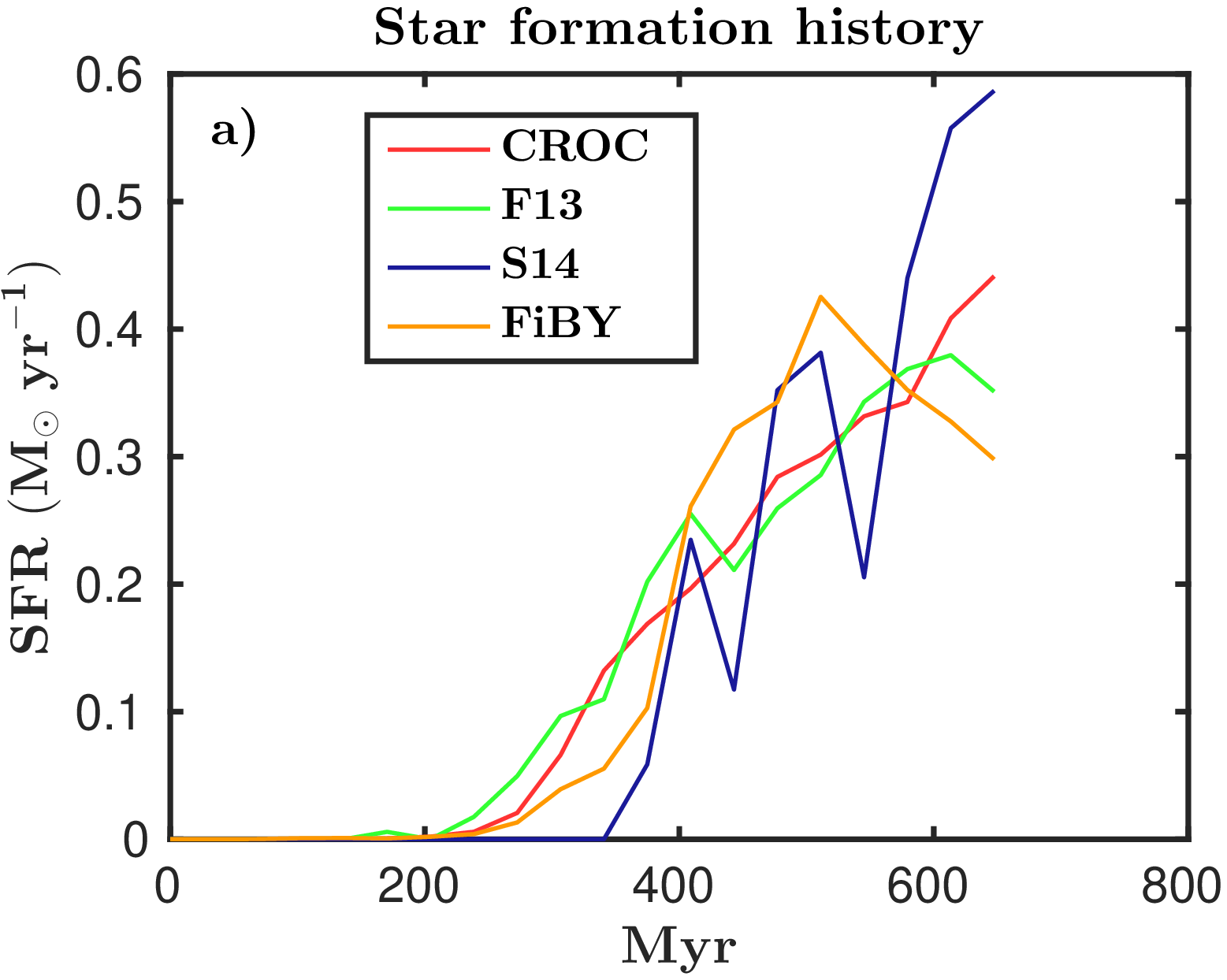}{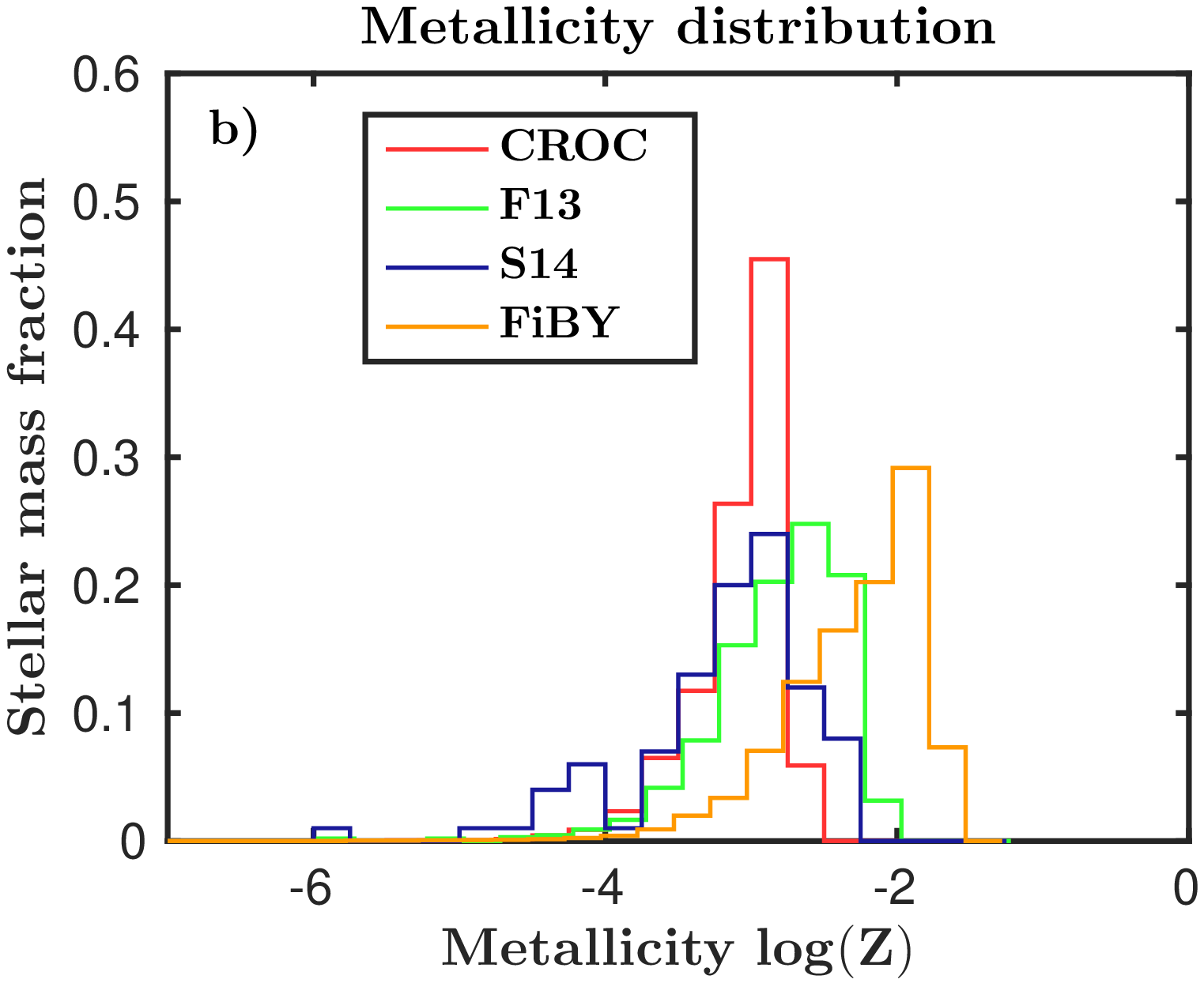}
\caption{Examples of the star formation histories and internal metallicity distributions for four galaxies with $M_\mathrm{stars}\sim 10^8\ M_\odot$ at $z=7$. The line colour represents the simulation suite from which each galaxy orignates: CROC (red), F13 (green), S14 (blue) ans FiBY (orange). {\bf a)} Star formation rate as a function of time up to the age of the Universe at $z=7$. All simulations predict a long-term trend of increasing star formation activity, but in some cases with substantial short-term fluctuations due to feedback and mergers. {\bf b)} The stellar metallicity distribution within the four simulated galaxies. In all cases, the mass-averaged metallicity $Z$ is below solar ($Z_\odot\approx 0.014$; \citealt{Asplund09}), although a significant fraction of stars have attained solar and even supersolar abundances in the FiBY galaxy shown.
\label{SFH_Z}}
\end{figure*}

In Figure~\ref{SFH_Z}, we show the internal metallicity distributions and star formation histories of one typical galaxy of mass $M_\mathrm{stars}\sim 10^8\ M_\odot$ at $z=7$ from each simulation suite. While fluctuations in star formation rates (SFRs) have occurred throughout the formation of these objects, the SFRs have on average been rising since the formation of their first stars \citep[as previously reported by e.g.][]{Finlator11,Jaacks12,Shimizu14}. This is exemplified in Figure~\ref{SFH_Z}a, where we plot the SFR as a function of time for these four simulated galaxies. For plotting purposes, we have here renormalized all four objects to exactly $M_\mathrm{stars}=10^8\ M_\odot$ to get rid of offsets due to slight differences in mass. Star formation rate fluctuations on time scales of 10--100 Myr of the type seen here (most notably in the S14 and FiBY cases) are produced by a combination of feedback effects and mergers and are commonly seen in simulations of high-redshift galaxies \citep[e.g.][]{Ma15,Kimm15}. 

All four suites consistently predict the metallicity distribution within a single $M_\mathrm{stars}\sim 10^8\ M_\odot$ galaxy at $z=7$ to be wide (varying by $\approx$ 2 dex) and to have a mass-weighted average $Z<Z_\odot$. Only a very small fraction of stars (always $< 10\%$ and typically much fewer) are predicted to be in the extremely metal-poor range ($Z<10^{-5}$) where the IMF is suspected to become top-heavy \citep[e.g.][]{Dopcke13,Safranek-Shrader14}. There are, however, substantial systematic differences in average metallicity between the simulations, as exemplified in Figure~\ref{SFH_Z}b -- with CROC typically predicting the lowest mean metallicties and FiBY the highest. At $M_\mathrm{stars}\sim 10^8 \ M_\odot$, the mean metallicity varies from $Z\approx 10^{-3}$ (i.e. $\sim 0.1\ Z_\odot$) to $Z\approx 10^{-2}$ (only slightly subsolar) between the simulation suites. However, these metallicity variations do not have any significant impacts on the results in this paper, as demonstrated in Section~\ref{diagnostic_nodust}.

\subsection{SED modelling}
\label{SED_modelling}
\subsubsection{Stellar and nebular contributions to the SED}
The internal stellar age and metallicity distribution from each simulated galaxy is stored as a collection of star particles (mass $\sim 10^3$--$10^6 \ M_\odot$ depending on simulation resolution), each considered to have a single age and metallicity. By using the Yggdrasil spectral synthesis code \citep{Zackrisson11} to attach a suitable model spectral energy distribution (SED) to each such star particle and summing over all star particles belonging to specific galaxy, the complete SED of that galaxy is generated. The end product is an SED in the rest-frame 0.1216-1 $\mu$m wavelength range (relevant for JWST/NIRSpec observations of $z>6$ galaxies) that takes both starlight, nebular emission (nebular emission lines plus nebular free-bound and free-free continuum), dust attenuation and LyC leakage into account. 

The first step of this procedure is to generate a stellar SED suitable for each star particle from a pre-generated grid of single stellar population (SSP) models covering a wide range in age $t$ and metallicity $Z$. Here, we have chosen to interpolate this grid in $\log(t)$ and $\log(Z)$ to produce this star particle SED. Tests with alternative computational schemes (interpolation in linear $t$ and $Z$ or by simply picking the closest grid SSP model) indicate that the end results are not particularly sensitive to how this interpolation is carried out. 

To evaluate the impact of different assumptions concerning stellar evolution, a number of alternative SSP grids have been used. The one we consider as the default option in this paper consists of Starburst99 \citep{Leitherer99} SSP stellar population spectra based on Geneva tracks with high mass-loss and no rotation at $Z=0.001$--0.040, extended to lower metallicities with \citet{Raiter10} SSP models at $Z=10^{-5}$ and $Z=10^{-7}$. All models have been scaled to match the \citet{Kroupa} universal stellar initial mass function (IMF). 

This assumption of an invariant IMF can certainly be questioned at metallicities close to the Population III regime ($Z\sim 10^{-5}$, $10^{-7}$ in our grid), where the IMF is expected to turn top-heavy. However, as discussed in section~\ref{simulations} and ~\ref{diagnostic_nodust}, this choice has a very small impact on the majority of galaxies in our simulations, since very few star particles tend to have metallicities in this range. 

As alternatives to our baseline grid, we also consider versions for which the Starburst99 Geneva models have been replaced by Starburst99 Padova-AGB models \citep{Vazquez&Leitherer} at $Z=0.0004$--0.050 or by BPASS v.2.0 models \citep{Stanway16} for binary stars at $Z=0.001$-0.030. 

For each stellar population SSP grid, nebular spectra for different LyC escape fractions $f_\mathrm{esc}$ have also been generated using the photoionization code Cloudy \citep{Ferland13}, based on the assumption of spherical, constant-density nebulae ($n(\mathrm{H})=100$ cm$^{-3}$) with the same metallicity as that of the stars. Within these nebulae, the electron temperature responds self-consistently to changes in the ionizing SED \citep[see][for further details]{Zackrisson11,Zackrisson13}. 

As discussed in \citet{Zackrisson13}, LyC leakage can take place through two basic scenarios -- leakage  through a radiation-bounded nebula with holes or leakage through a density-bounded nebula. As demonstrated in that paper, the two mechanisms give rise to similar predictions in the EW(H$\beta$)-$\beta$ diagram in the dust-free case. Things get more complicated when dust effects are considered, since the direct absorption of Lyman continuum photons of dust (see section~\ref{LyC_extinction}) is likely to be more important in the density-bounded case. In this paper, we for simplicity focus on the case of radiation-bounded nebulae with holes free of both gas and dust. In this case, the LyC escape fraction is simply regulated by the gas covering fraction $f_\mathrm{cov}$ adopted in Cloudy ($f_\mathrm{esc}=1-f_\mathrm{cov}$). Since a number of recent observations (see section~\ref{density-bounded}) indicate that density-bounded nebula may also be highly relevant, we are planning a follow-up paper dedicated to this specific mode of LyC escape.

\begin{figure}[t]
\centering
\includegraphics{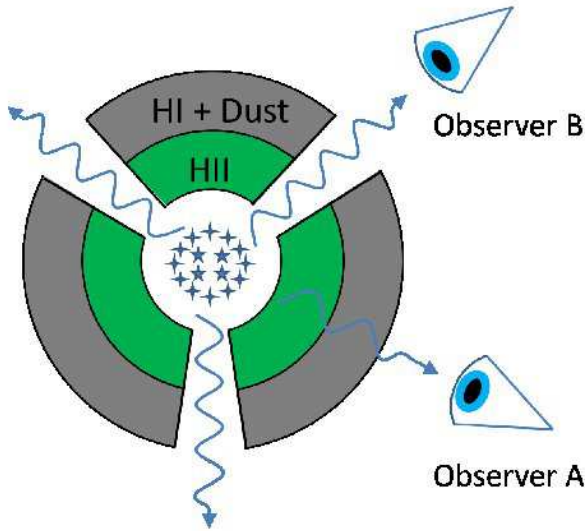}
\caption{Schematic illustration of the two different viewing angles considered when simulating the impact of dust attenuation on observed galaxy SEDs. The galaxy is here assumed to consist of a central star-forming region, a surrounding photoionized region (green) and an outer shell of neutral gas and dust (grey). Observer A happens to be viewing the galaxy from an angle which does not include any direct starlight. Since observer A receives nebular emission from the whole photoionized region, the relative impact of nebular emission on the galaxy SED observed by A drops with increasing LyC escape fraction $f_\mathrm{esc}$. However, the impact of dust attenuation on this SED does not necessarily weaken with increasing $f_\mathrm{esc}$ since all the light received by A passes through dusty regions (grey). The galaxy SED seen by observer B, on the other hand, contains a fraction of direct, unattenuated starlight.   
\label{schematic}}
\end{figure}

\subsubsection{Dust attenuation}
To account for the effects of dust in the rest-frame UV/optical, we consider two alternative recipes developed by F13 and  S14 for predicting the attenuation of simulated galaxies. In the F13 recipe, the rest-frame color excess $E(B-V)$ is based on a deterministic part related to the overall stellar population metallicity $Z$, plus a random component $\delta E(B-V)$:
\begin{equation}
E(B-V)=9.0Z^{0.9} + \delta E,
\end{equation} 
where $\delta E$ is a Gaussian component with a variance equal to half of the metallicity-dependent component. In the case of our $z=7$ sample of galaxies from the \citet{Finlator13} simulations, this gives an average $E(B-V)\approx 0.026$, which converts into average $\langle A(V) \rangle\approx 0.1$--0.2 magnitudes (where the exact value depends on the attenuation law; see below).

The S14 recipe is based on the metallicity, total stellar mass and the spatial distribution of stars within each  galaxy, with free model parameters calibrated against the rest-frame UV luminosity function at $z=7$ \citep[for details, see][]{Shimizu14}. The end result is a prediction of the 1500 \AA{} attenuation ($A(1500)$) for each galaxy, which converts into an average optical extinction that is somewhat higher than for the F13 case, with $\langle A(V)\rangle \approx 0.2$--0.4 depending on the attenuation law.

\begin{figure*}
\centering
\plottwo{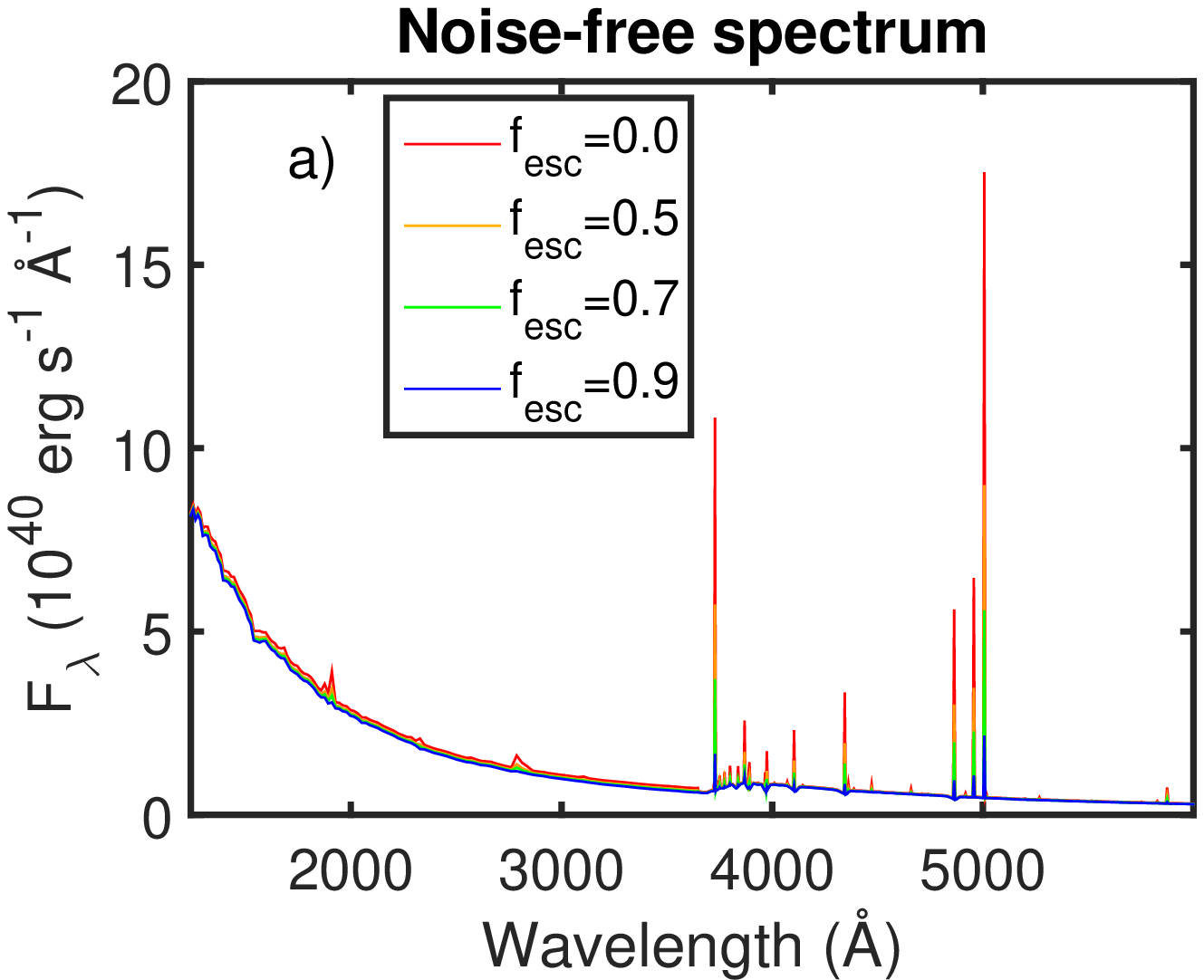}{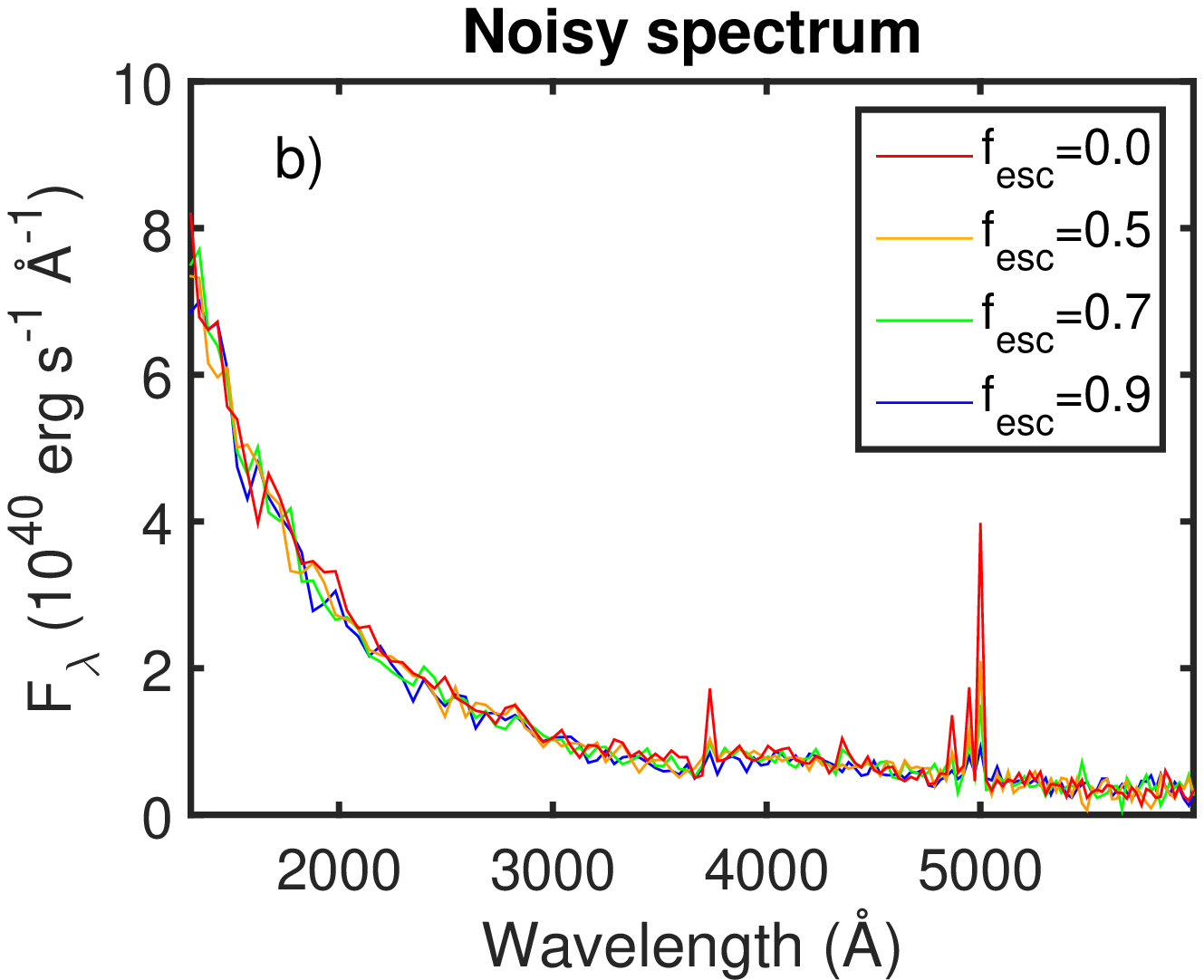}
\caption{Synthetic spectra of a single $M_\mathrm{stars}\approx 7\times10^8\ M_\odot$, $m_\mathrm{AB}\approx 27$, dust-free \citet{Shimizu14} galaxy for different $f_\mathrm{esc}$: 0.0 (red), 0.3 (orange), 0.5 (green) and 0.7 (blue).  {\bf a)} Noiseless spectra {\bf b)} Spectra degraded to the spectral resolution of the JWST/NIRSpec $R=100$ prism, and with noise level corresponding to a 10 h exposure. The $f_\mathrm{esc}$ parameter mainly affects the strength of the emission lines relative to the continuum, and this effect remains detectable for the strongest emission lines (here [OII]$\lambda$3727, H$\beta$, [OIII]$\lambda$4959 and [OIII]$\lambda$5007) even after the introduction of observational noise. In Figure~\ref{SEDs_zoomin}, we also show a zoom-in on the region around the latter three lines.\label{SEDs}}
\end{figure*}

Once either $E(B-V)$ -- the Finlator case -- or $A(1500)$ -- the Shimizu case -- has been established, we consider a number of different wavelength-dependent dust attenuation laws to apply reddening throughout the full wavelength range of our model SEDs: the LMC and SMC attenuation laws by \citet{Pei92} or the \citet{Calzetti00} attenuation law for starburst galaxies.  In the case of the LMC and SMC attenuation laws, we assume the attenuation to affect the stellar and nebular portions of the SED by an equal factor. In the \citet{Calzetti00} case, we allow the reddening of the stellar component either to be equal to that of the nebular one ($E(B-V)_\mathrm{stars}=E(B-V)_\mathrm{neb}$), or a factor of 0.44 lower ($E(B-V)_\mathrm{stars}=0.44E(B-V)_\mathrm{neb}$, i.e. the case actually advocated by \citealt{Calzetti00}). The higher $E(B-V)$ implied for the nebular component in the latter case is usually interpreted as an effect of age-dependent attenuation, in the sense that young star-forming regions are embedded within thick dust cocoons that disperse over time \citep[][]{Charlot & Fall, Bergvall16}. One would then expect the $f=E(B-V)_\mathrm{neb}/E(B-V)_\mathrm{stars}$ ratio to depend on star formation history, effectively approaching unity in cases where all stars have the same age or when young star-forming regions come to dominate the spectrum. Indeed, observational studies support this, indicating that $f$ depends on the specific star formation rate \citep[e.g.][]{Price14,Puglisi16}. To capture this variation in $f$ across a sample of simulated galaxies, it would seem more realistic to apply age-dependent attenuation corrections to each star particle separately, rather than relying on galaxy-wide attenuation corrections. However, tests using the age-dependent dust attenuation recipe developed by \citet{Bergvall16} -- with parameters tuned to make the average UV slope $\beta$ of the $z=7$ galaxy population agree with observations -- applied to each star particle indicates that this produces results that are qualitatively similar to the $E(B-V)_\mathrm{stars}=0.44E(B-V)_\mathrm{neb}$ case in the EW(H$\beta$)-$\beta$ diagram. Throughout the rest of this paper, we will therefore consider galaxy-wide attenuation corrections only.

The effect of dust reddening on the observed SED may potentially be reduced in the case of very high $f_\mathrm{esc}$, but the impact of this effect is likely to depend on the geometry of gas, stars and dust, the level of leakage isotropy and the viewing angle of the observer. This is schematically illustrated in Figure~\ref{schematic}, where two observers A and B are viewing an ionization-bounded nebula with holes from different angles. Observer A measures an SED where all the starlight and nebular emission has passed through a dust screen, which -- at least to first order -- may be considered to be unaffected by $f_\mathrm{esc}$ (this is hereafter referred to as scenario A). Observer B, on the other hand, measures an SED which contains a fraction of unattenuated starlight. In the case of highly isotropic leakage through many thin escape channels, a fraction $f_\mathrm{esc}$ of the starlight directed towards observer B may be unaffected by dust whereas a fraction $1-f_\mathrm{esc}$ experiences dust attenuation (hereafter scenario B). However, these are admittedly not the only options, since anisotropic leakage could allow a fraction of unattenuated starlight different from the global $f_\mathrm{esc}$ to escape in the direction of B. Throughout this paper, we consider scenario A the default option, but also explore the consequences of scenario B on the EW(H$\beta$)-$\beta$ diagram. 

\subsubsection{Simulating JWST/NIRSpec data quality}
To provide realistic simulations of galaxy spectra as they would appear when observed with the JWST/NIRSpec instrument, our model also allows the option of degrading the quality of the simulated SEDs by taking the noise and limited spectral resolution into account, following the JWST/NIRSpec-specific recipe described in \citet{Jensen et al.}. 

As an example of how $f_\mathrm{esc}$ affects the SED in the wavelength range relevant for JWST/NIRSpec, we in Figure~\ref{SEDs} show the predicted SED of a dust-free, $M_\mathrm{stars}\approx 7\times 10^8\ M_\odot$ galaxy from the \citet{Shimizu14} batch at $z=7$ for various $f_\mathrm{esc}$, with and without instrumental effects (noise and degraded spectral resolution). At this redshift, this object attains an apparent magnitude of $m_\mathrm{AB}\approx 27$ in the rest-frame ultraviolet, where the SED is continuum dominated.

Here, we have assumed a very long exposure time ($t_\mathrm{exp}=10$ h) using the $R=100$ NIRSpec prism. As argued by \citet{Zackrisson13}, $f_\mathrm{esc}$ mainly affects the strengths of emission lines compared to the continuum, and while observational effects will render the weaker emission lines impossible to detect, several strong emission lines remain potentially useful as LyC leakage diagnostics, including [OII]$\lambda$3727, H$\gamma$, H$\beta$, [OIII]$\lambda$4959 and [OIII]$\lambda$5007. While the [OII]$\lambda$3727 and [OIII]$\lambda$5007 emission lines actually appear as the strongest lines in the SED, the relative strengths of these lines are sensitive to the mechanism of LyC leakage, and may be very different in case of a density-bounded nebula compared to the matter-bounded nebula with holes assumed here. Following \citet{Zackrisson13}, we hence focus on the diagnostic properties of the Balmer line H$\beta$ (4863 \AA{}; the third strongest line in the SED plotted), as the predictions for this line are very similar in the two cases. The H$\beta$ line is also convenient because the physics of hydrogen recombination is relatively simple at densities typical of the interstellar medium in galaxies -- in the absence of LyC leakage, the H$\beta$ luminosity is not significantly affected by the gas density, the ionization parameter or the gas metallicity. Hence, there is no significant difference between models where the whole stellar population is considered to be a central light source surrounded by nebular gas (as assumed here and illustrated in Figure~\ref{schematic}), or whether the stars and gas are mixed. This is in stark contrast to the behaviour of
other potentially useful lines like CIII]$\lambda$1909, CIV$\lambda$1549, [OII]$\lambda$3727, [OIII]$\lambda$4959 and [OIII]$\lambda$5007, which are more sensitive to the physical state of the nebula \citep[e.g.][]{Nakajima & Ouchi,Stasinska et al.,Jaskot & Ravindranath}. 

\begin{figure}[t]
\centering
\plotone{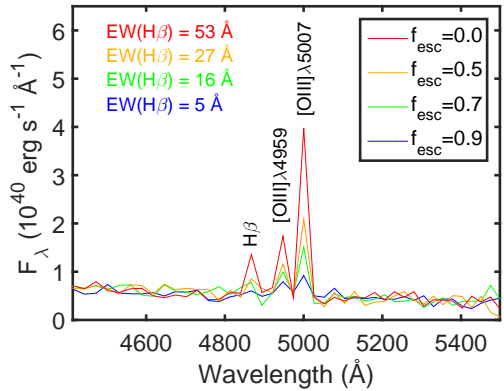}
\caption{The rest-frame wavelength interval around H$\beta$, [OIII]$\lambda$4959 and [OIII]$\lambda$5007 extracted from the noisy, \citet{Shimizu14} SEDs ($m_\mathrm{AB}\approx 27$, $t_\mathrm{exp}=10$ h, $R=100$) of Fig.~\ref{SEDs}b. The rest-frame EW(H$\beta$) for this particular galaxy are listed for $f_\mathrm{esc}$: 0.0 (red), 0.5 (orange), 0.7 (green) and 0.9 (blue). As seen, low EW(H$\beta$) serves as a tell-tale signature of high-$f_\mathrm{esc}$, and although a robust detection of the H$\beta$ emission line will be very challenging at $f_\mathrm{esc}>0.5$, an upper limit on EW(H$\beta$) can still be imposed as long as the expected position of H$\beta$ is known from the redshift measured from the stronger [OIII]$\lambda$4959 and [OIII]$\lambda$5007 lines.   
\label{SEDs_zoomin}}
\end{figure}

In Fig.~\ref{SEDs_zoomin}, we present a zoom-in of the SEDs from Fig.~\ref{SEDs}b to illustrate the expected data quality in the wavelength range surrounding the $H\beta$ line. As seen, the H$\beta$ line becomes challenging to detect at $f_\mathrm{esc}\gtrsim 0.5$, but as long as its expected position can be inferred from the wavelengths of the stronger [OIII]$\lambda$4959 and [OIII]$\lambda$5007 lines, it should still be able to set an upper limit on EW(H$\beta$), which is quite sufficient to single out galaxies with very high levels of LyC leakage and set a lower lower limit on $f_\mathrm{esc}$. At $f_\mathrm{esc}\gtrsim 0.9$, however, all of these lines become difficult to detect, which makes for substantial uncertainties in determining the redshift, unless this can be done from the detection of the Lyman-$\alpha$ break \citep[as in][]{Oesch16}.

\section{The EW(H$\beta$)-$\beta$ diagram}
\label{diagnostic}
\subsection{The case without dust attenuation}
\label{diagnostic_nodust}

\begin{figure*}[t]
\centering
\includegraphics[scale=0.5]{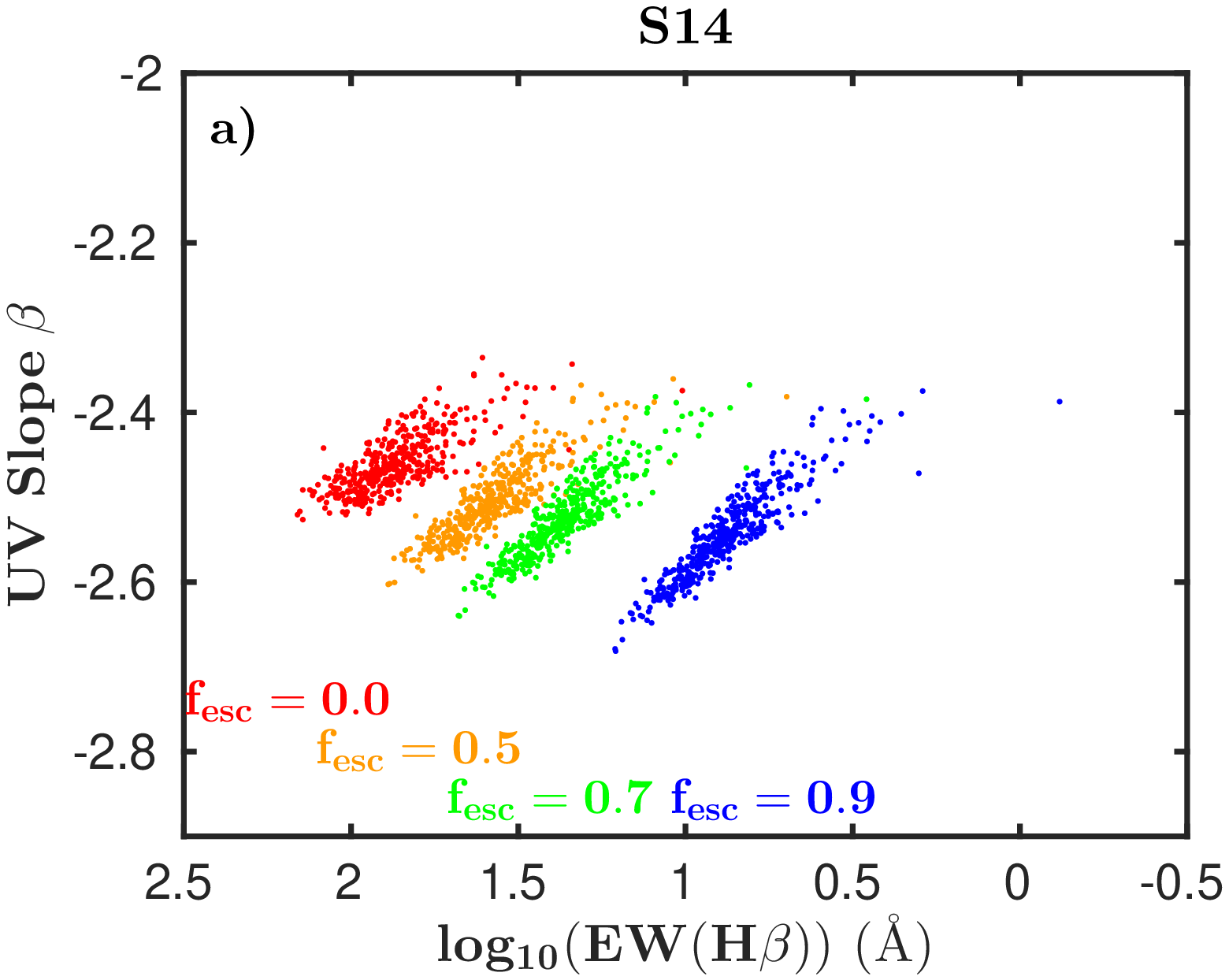}\includegraphics[scale=0.5]{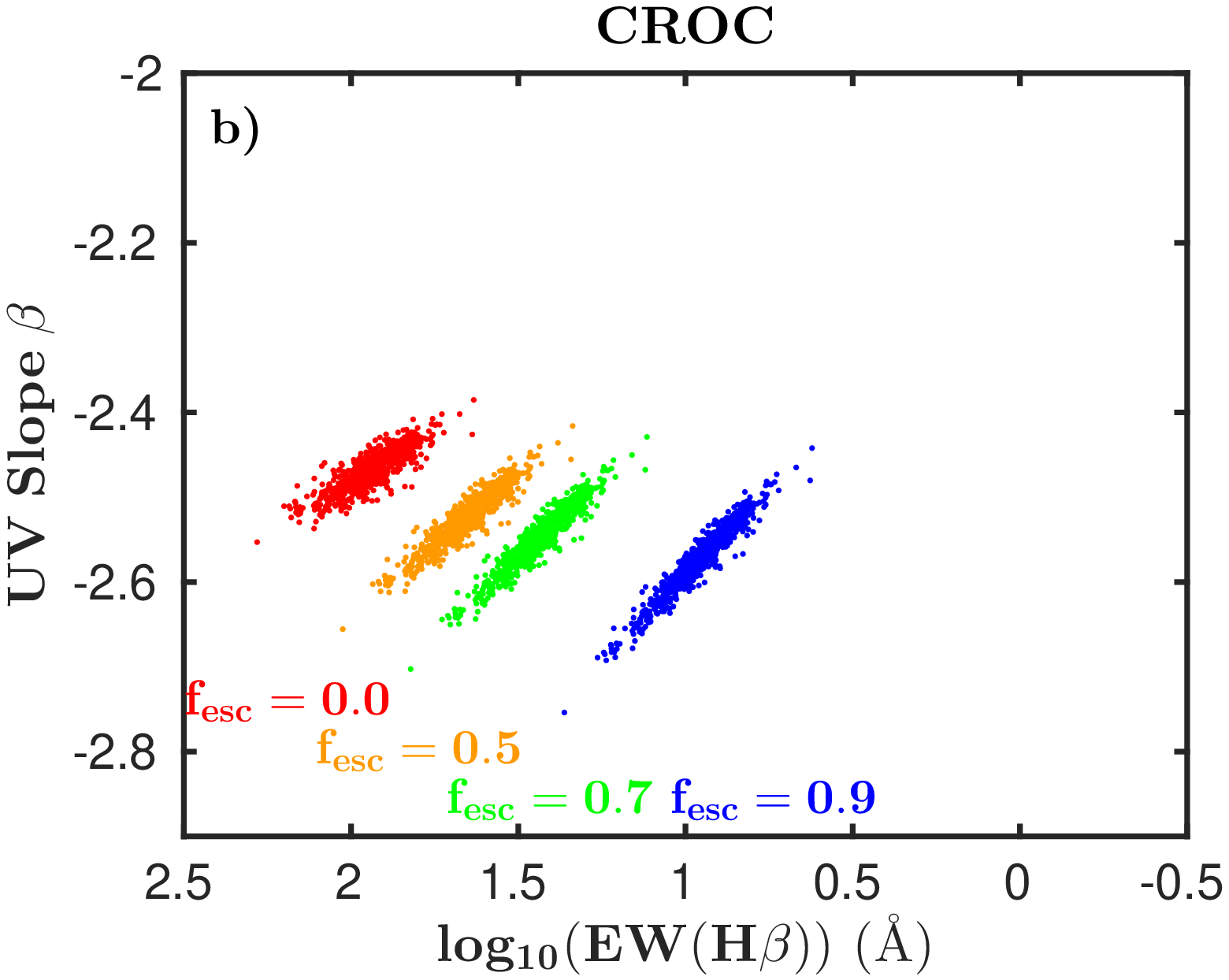}\\
\includegraphics[scale=0.5]{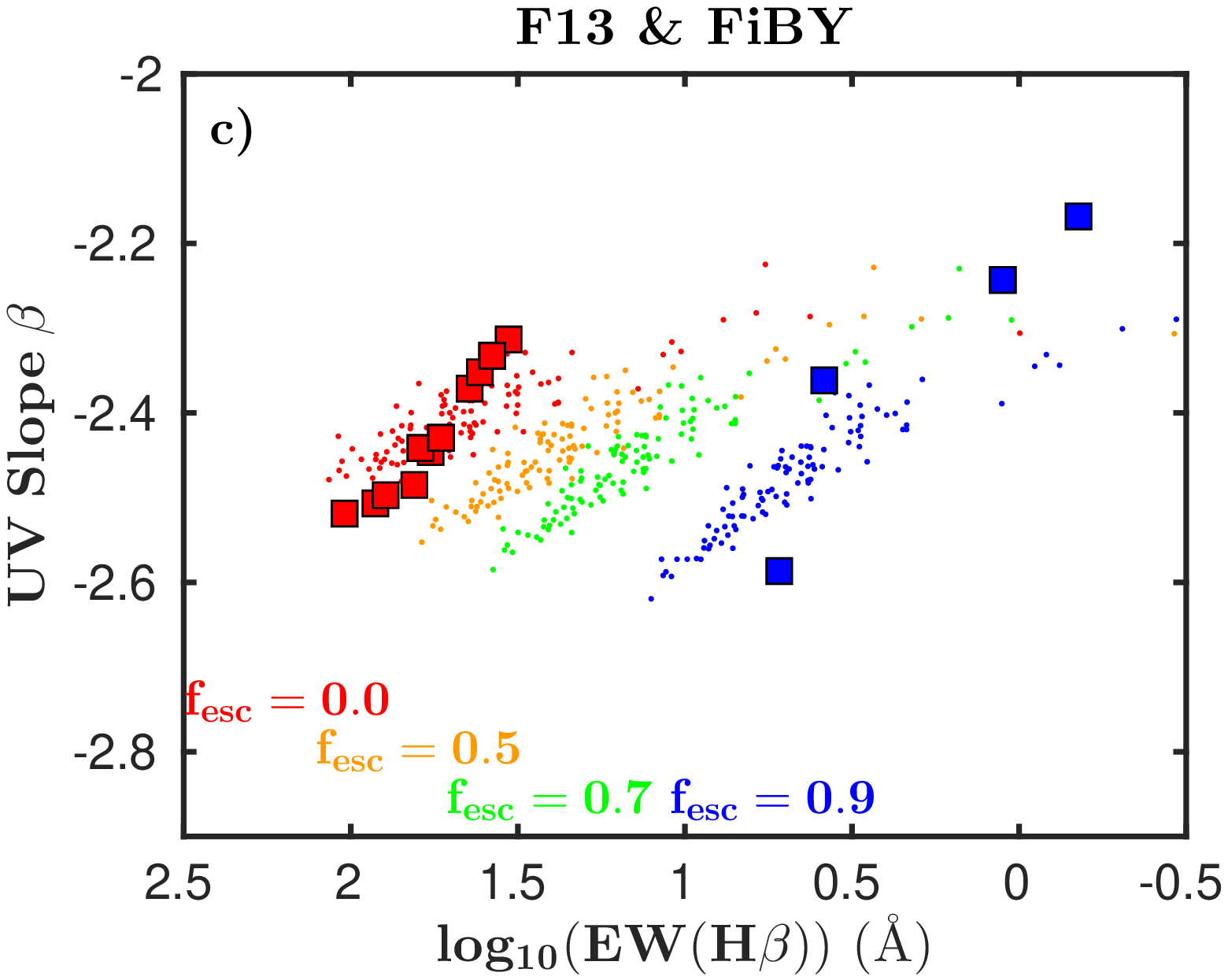}\includegraphics[scale=0.5]{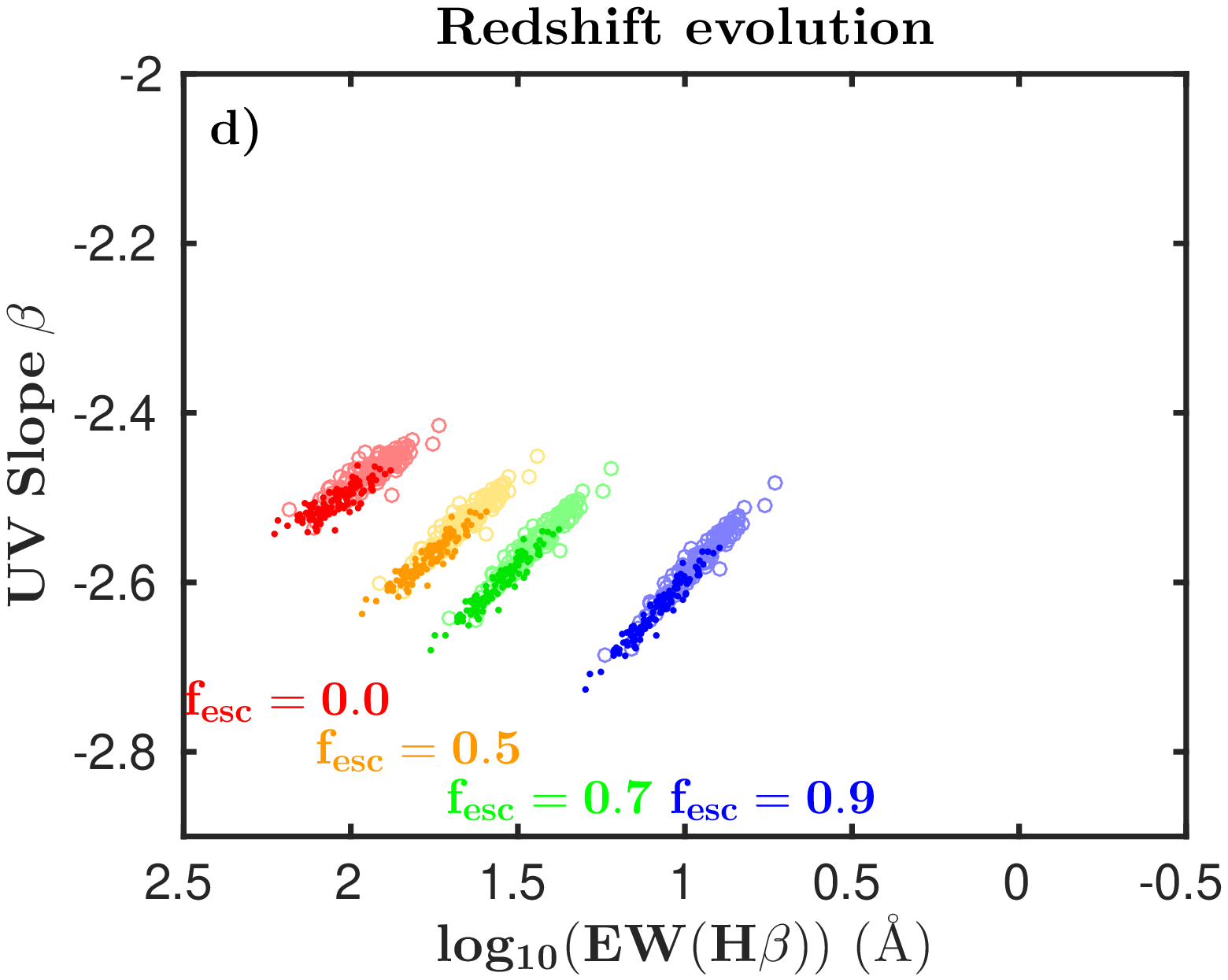}\\
\caption{The position of synthetic, dust-free galaxies in the EW(H$\beta$)-$\beta$ diagram for escape fractions $f_\mathrm{esc}=0.9$ (blue markers), 0.7 (green), 0.5 (orange) and 0.0 (red). {\bf a)} Galaxies at $z=7$ from \citet{Shimizu14}; {\bf b)} Galaxies at $z=7$ from CROC; {\bf c)} Galaxies at $z=7$ from \citet{Finlator13} and at $z=6$--8 from the FiBY simulations (squares). Please note that the LyC escape fractions in the \citet{Shimizu14}, CROC and \citet{Finlator13} simulations are {\it assumed}, whereas they are {\it predicted } in the case of FiBY, and colored according to their approximate $f_\mathrm{esc}$. 
{\bf d)} Comparison of the positions of 100 randomly selected CROC galaxies with $M_\mathrm{stars} \geq 10^7\ M_\odot$ at $z=9$ (dark dots) and $z=7$ (light circles). Only modest evolution in the EW(H$\beta$) and $\beta$ distributions is seen between these redshifts.
\label{no_dust}}
\end{figure*}

In Figure~\ref{no_dust}, we show the behavior of dust-free galaxies with various $f_\mathrm{esc}$ from the S14, CROC, F13 and FiBY simulation suites in the EW(H$\beta$)-$\beta$ diagram. Here, $\beta$ (defined as $f_\lambda\propto \lambda^\beta$) is derived from the ten wavelength intervals in the rest-frame range $\approx 1270$–-2580 \AA{} defined by \citet{Calzetti94}. All SEDs are here based on Geneva stellar evolutionary tracks. In the case of the first three simulation suites, we explore the effects of $f_\mathrm{esc}=0$ (red dots), 0.5 (orange), 0.7 (green) and 0.9 (blue), whereas $f_\mathrm{esc}$ is self-consistently predicted in the case of the FiBY galaxies. The latter objects have been color-coded to match the $f_\mathrm{esc}$ bins that they are deemed to fall closest to (most at $f_\mathrm{esc}=0$ but a few at  $f_\mathrm{esc}=0.9$).

While galaxy-to-galaxy variations in star formation history and metallicity distributions induces a considerable scatter in EW(H$\beta$)-$\beta$ space, the different $f_\mathrm{esc}$ models still produce relatively well-separated regions in this diagram. This suggests that the method proposed by \citet{Zackrisson13} to identify high-$f_\mathrm{esc}$ galaxies is relatively robust to uncertainties regarding the ages and metallicity stellar populations of galaxies at these redshifts, at least in the absence of dust attenuation. In Figure~\ref{no_dust}d, we compare the predicted distribution of objects in the EW(H$\beta$)-$\beta$ diagram for 100 randomly-selected CROC galaxies at $z=7$ and $z=9$. While $z=9$ galaxies have slightly bluer $\beta$ and higher EW(H$\beta$) due to their somewhat lower ages, the evolution between these redshifts is marginal, which means that our $z=7$ results should apply to the whole $z=7$--9 interval.

The regions spanned by different $f_\mathrm{esc}$ are also relatively consistent among the four simulations used. However, since the simulations display different galaxy-to-galaxy diversity in star formation history and metallicity distribution, the compactness of the regions differ. For instance, the CROC galaxies exhibit smaller and more well-defined regions than the others, mainly due to their smoother star formation histories.

\subsection{Uncertainties related to stellar evolution}
\label{stellar_evolution}
While the diagnostic value of the EW(H$\beta$)-$\beta$ diagram in identifying high-$f_\mathrm{esc}$ galaxies appears relatively robust to variations in star formation history and metallicity (at least at the level of variation seen in the four simulation suites explored), there are other model uncertainties that could potentially bias the outcome. Any stellar population property that substantially boosts the ionizing stellar ultraviolet continuum compared to the non-ionizing ultraviolet continuum flux would  result in a bias of the $f_\mathrm{esc}$ inferred from observations, if not accounted for in the models. 

\begin{figure*}[t]
\centering
\plottwo{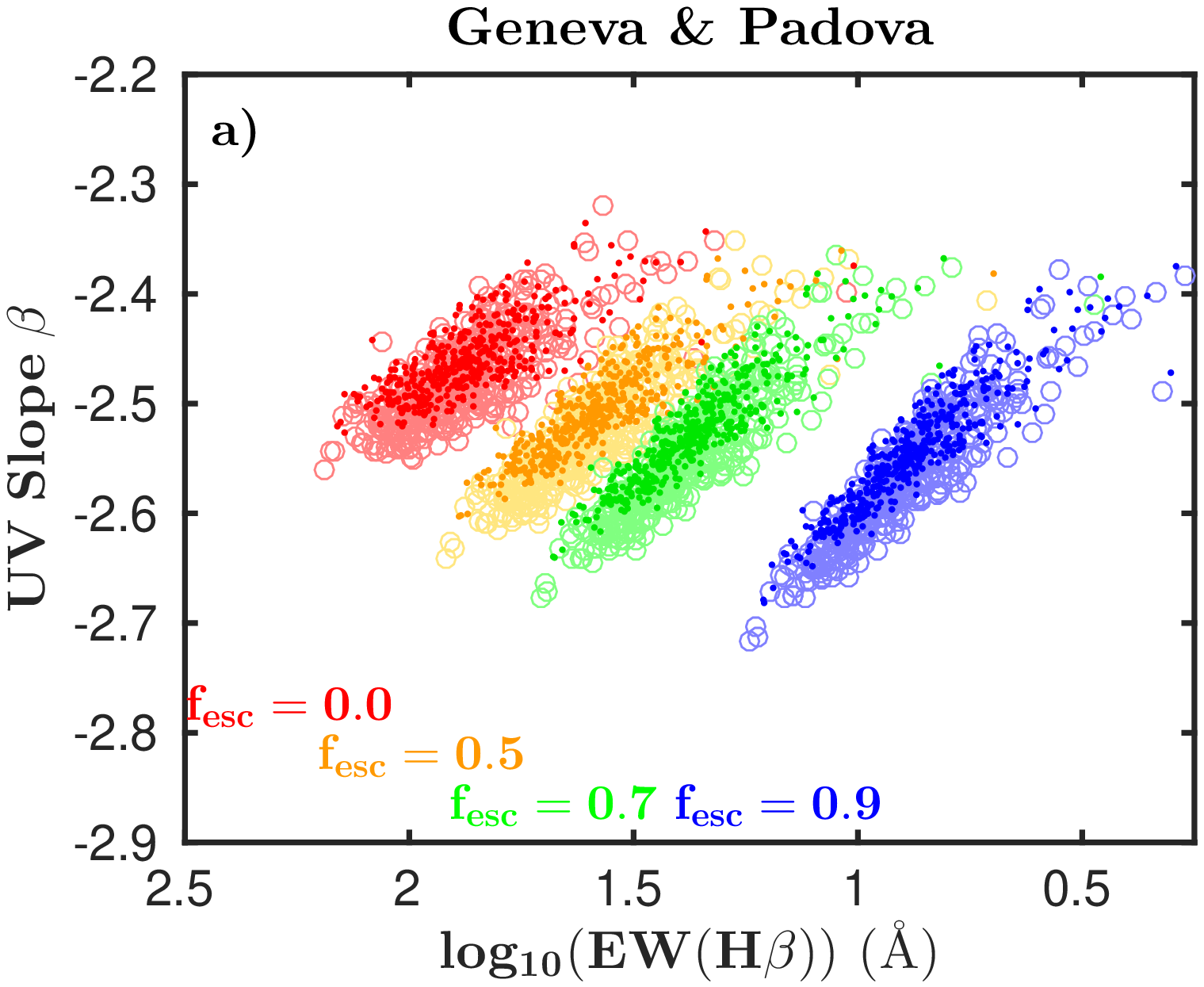}{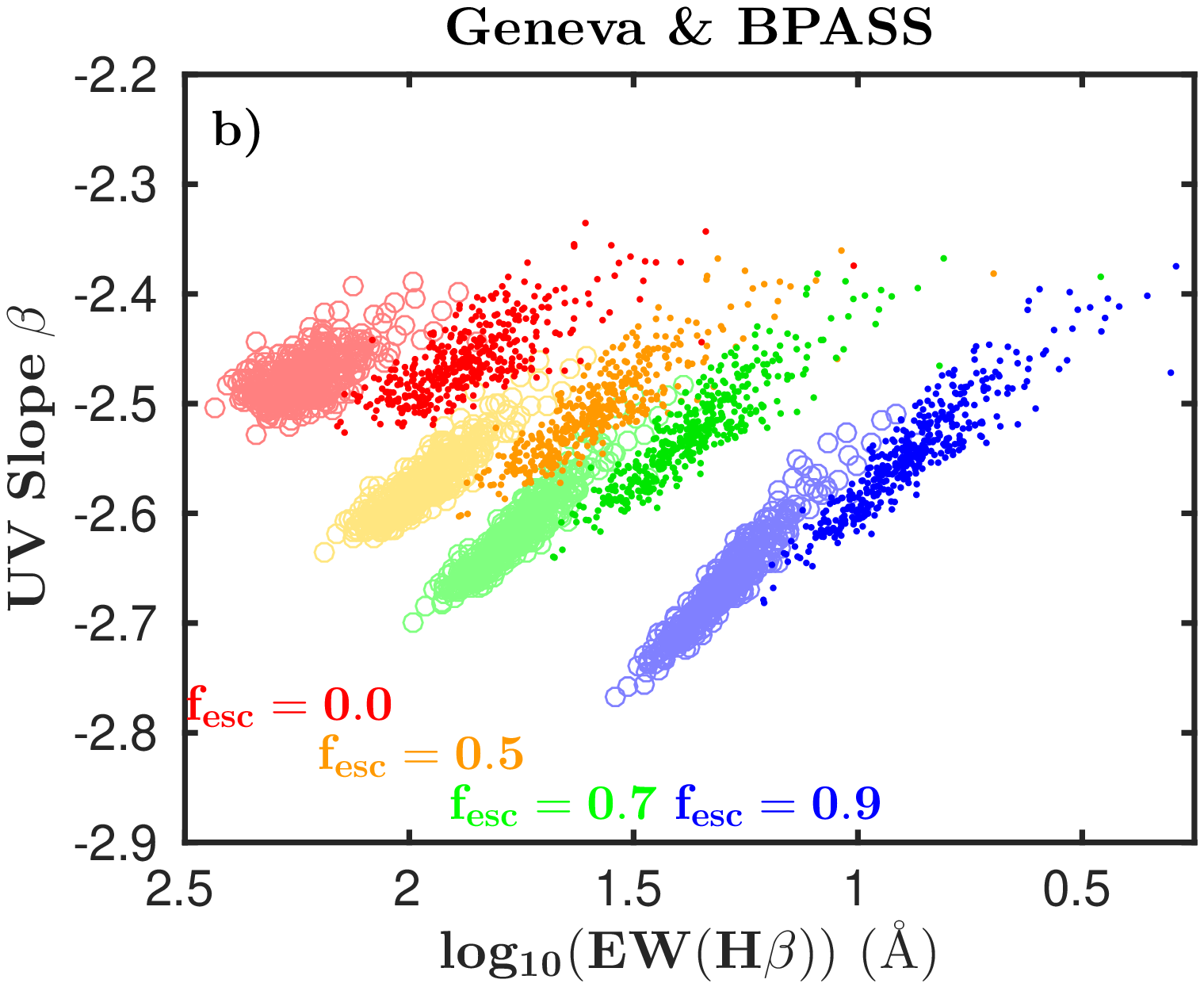}
\caption{The positions of dust-free galaxies at $z=7$ from the \citet{Shimizu14} simulations in the EW(H$\beta$)-$\beta$ diagram, under different assumptions concerning stellar evolution. The different colors represent LyC escape fractions $f_\mathrm{esc}=0.9$ (blue), 0.7 (green), 0.5 (orange) and 0.0 (red). {\bf a)} SEDs based on SB99 Geneva stellar evolutionary tracks (dark dots) compared to SB99 Padova stellar evolutionary tracks (light circles); {\bf b)} SEDs based on SB99 Geneva stellar evolutionary tracks (dark dots) compared to BPASS stellar models with binary evolution (light circles). While the difference in predictions based on Geneva and Padova tracks are minor, binary evolution (as implemented in the BPASS models) leads to substantial differences, which will inevitably introduce a bias in the $f_\mathrm{esc}$ inferred from observational data unless accounted for in the models. However, it should be possible to use the overall distribution of galaxies in the EW(H$\beta$)-$\beta$ diagram to test for model deficiencies of this type.\label{model_tracks}}
\end{figure*}

A generic prediction when approaching the population III metallicity regime ($Z\lesssim 10^{-5}$) is a boost in the characteristic mass of stars \citep[e.g.][]{Dopcke13,Safranek-Shrader14}. As demonstrated by \citet{Zackrisson13}, this would have a very pronounced impact on the EW(H$\beta$)-$\beta$ diagram by pushing objects upwards (towards redder $\beta$ slopes) through a
boost in the relative impact of nebular emission at fixed $f_\mathrm{esc}$. While our simulations predict mean metallicites $Z>10^{-4}$ for all galaxies sufficiently massive to be within range of NIRSpec at $z>6$, there is a minority of star particles with metallicities ranging from $Z=0$ to $Z\sim 10^{-5}$ in these objects, for which the choice of the \citet{Kroupa} IMF could certainly be questioned. However, for this to have any significant effect on the total SED also requires that these galaxies are caught in the phase where these particular star particles are younger than a few Myr. This condition is never met by more than $\approx 1\%$ of the star particles in any of our simulated galaxies and there is consequently no outliers predicted at very red $\beta$ and very high EW(H$\beta$) in Figure~\ref{no_dust}, even if the extremely top-heavy Population III IMF considered by \citet{Zackrisson13} is applied to the $Z< 10^{-5}$ particles. We therefore conclude that the assumptions on the behavior of the IMF at very low metallicities is not a major hurdle for our method to measure $f_\mathrm{esc}$. 

Uncertainties in the treatment of stellar evolution at higher metallicities can also be important, and recent studies have highlighted that both stellar rotation \citep[e.g.][]{Levesque12,Leitherer14,Topping15} and binary evolution \citep[e.g.][]{Eldridge09,Stanway16,Wilkins16,Ma16} may affect the ionizing fluxes of stellar populations by factors of a few. Since model grids that treat such effects have yet to be released for the full range of metallicities seen in our simulated galaxies, a completely self-consistent test of such effects is beyond the scope of the present study. However, since the fraction of star particles at metallicities $Z<10^{-4}$ is low, an assessment of the likely effects can nonetheless be made by exchanging the model grids at higher metallicities for those based on assumption different from those in our default batch. In Figure~\ref{model_tracks}a, we demonstrate the effects of switching from SB99 models based on Geneva stellar evolutionary tracks to Padova stellar evolutionary tracks in the case of the S14 galaxies. This turns out to have a very modest impact on the predicted position of galaxies in the EW(H$\beta$)-$\beta$ diagram, and would not bias $f_\mathrm{esc}$ estimates in any significant way. A more dramatic effect is seen in Figure~\ref{model_tracks}b, where we show the effect of making a corresponding switch from SB99 Geneva models for single stars to BPASS models for binary stars. In this case, both the EW(H$\beta$) and $\beta$ distributions of the galaxy population are shifted, by as much as $\Delta \log_{10}\mathrm{EW(H}\beta)\approx 0.5$ (i.e. a factor of $\approx 3$ shift in EW(H$\beta$)) and by $\Delta \beta\approx 0.15$. 

However, it should be possible to probe the observational data for signs of mismatches between the stellar evolution assumptions adopted in the simulations and the stellar evolution taking place in real galaxies, and thereby calibrate the models before attempting to infer $f_\mathrm{esc}$. Such calibrations can either be done at low redshifts, by simply testing models based on different assumptions concerning stellar rotation and binarity against local stellar populations \citep[e.g.][]{Wofford16}, or directly from JWST/NIRSpec observations of galaxies at $z>6$. As seen in Figure~\ref{model_tracks}b, because of the higher ionizing fluxes produced once binary stars are considered, BPASS predicts the presence of galaxies with $\log_{10} \mathrm{EW(H}\beta)>2$ galaxies at $z=7$ that are not predicted to exist in the case of models without binary evolution. Hence, the detection of a substantial number of objects with such properties would clearly invalidate our default models based on the evolution of single stars.

\subsection{Dust effects}
\label{dust}
At low levels of LyC leakage, the different dust attenuation recipes described in Section~\ref{SED_modelling} give rise to mean optical (rest-frame $V$-band) attenuation values of $A_V\approx 0.15$--0.4 mag, depending on the dust attenuation curve adopted, which leads to average UV $\beta$ slopes of $\beta \approx -2.2$ to $-2.0$, in rough agreement with current observations of $z\approx 7$ galaxies \citep[e.g.][]{Bouwens14}. 

\begin{figure*}
\centering
\includegraphics[scale=0.5]{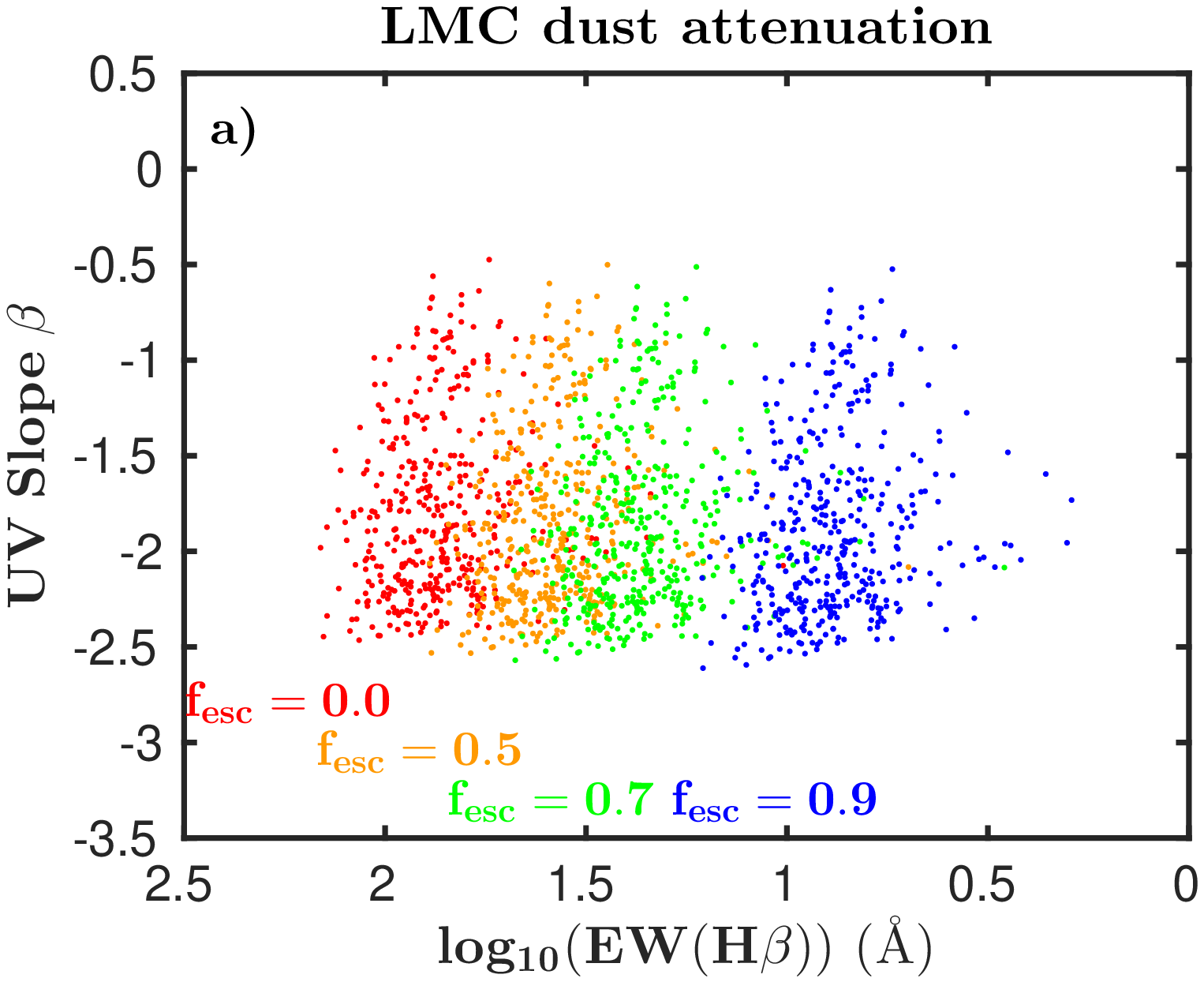}\includegraphics[scale=0.5]{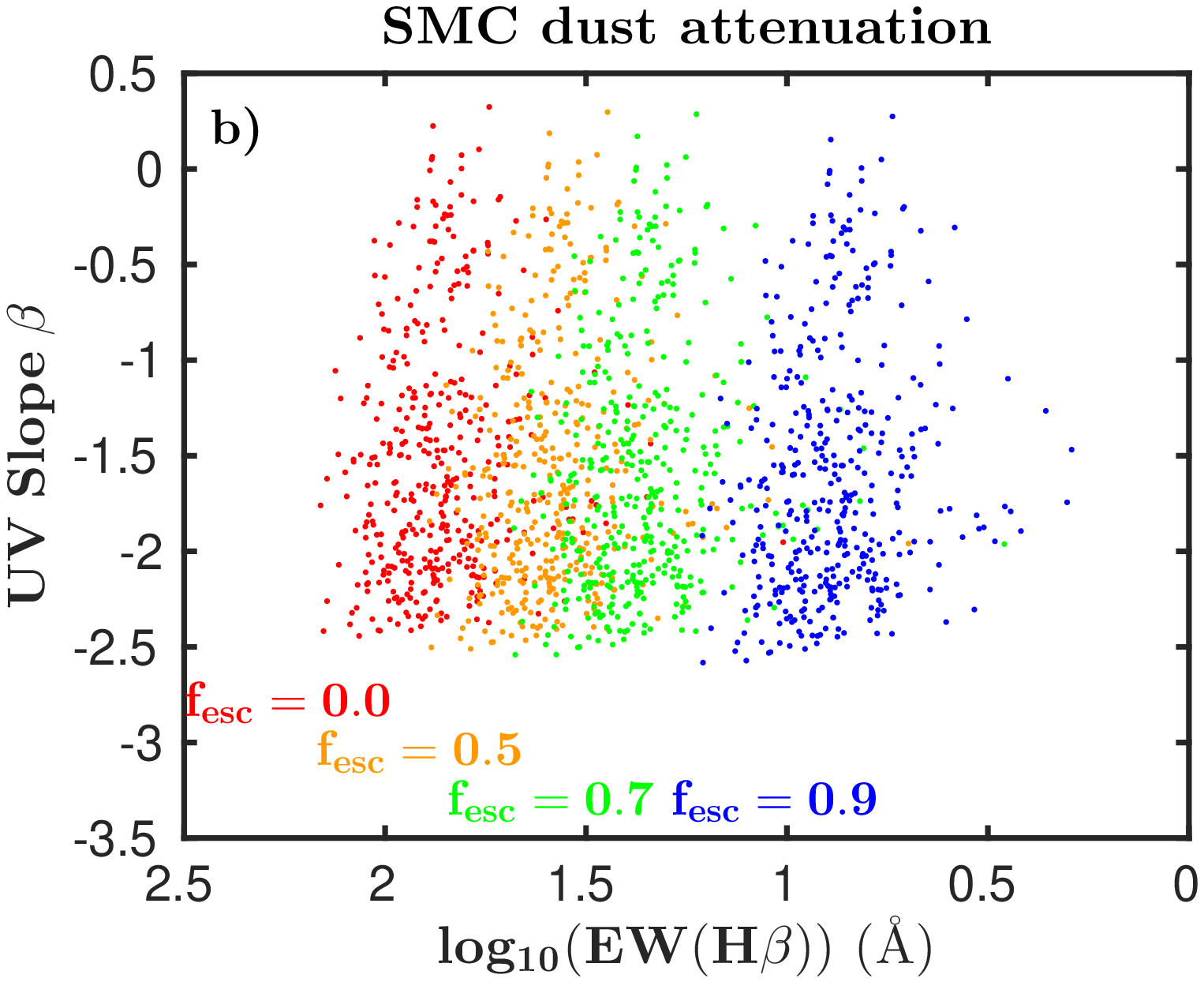}\\
\includegraphics[scale=0.5]{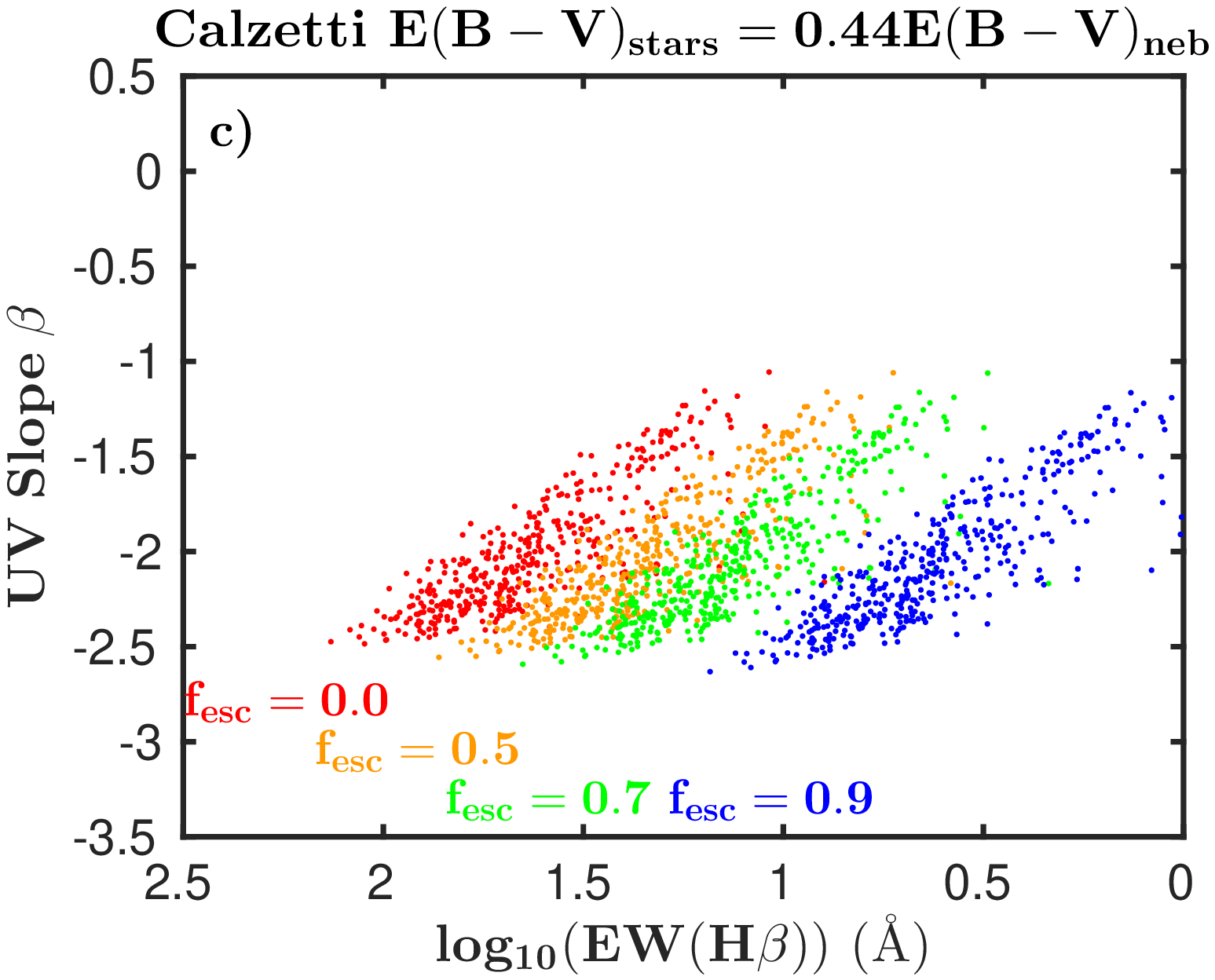}\includegraphics[scale=0.5]{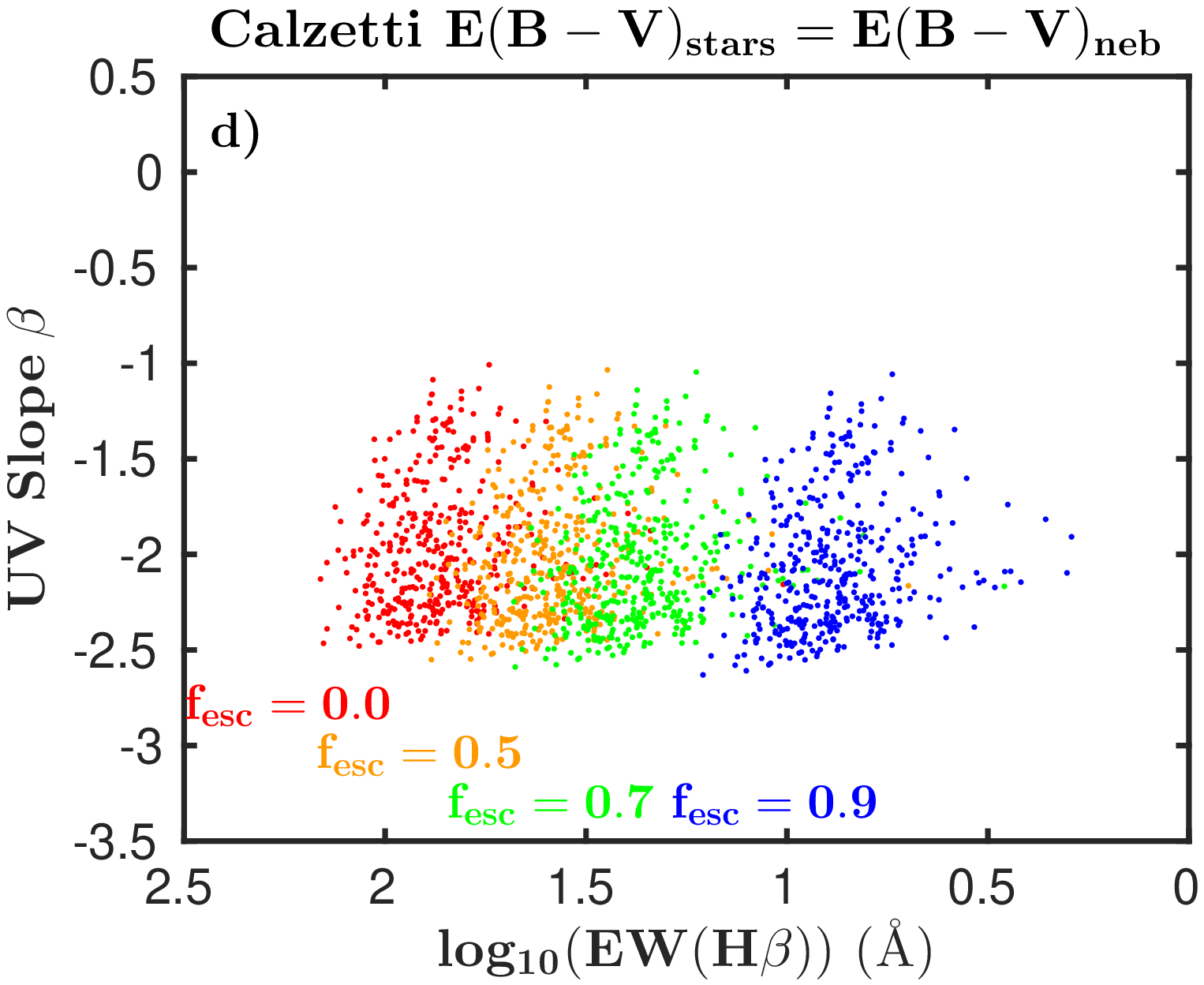}\\
\caption{Effects of dust attenuation on $M_\mathrm{stars}\geq 5\times 10^8\ M_\odot$ galaxies at $z=7$ from the \citet{Shimizu14} simulations in the EW(H$\beta$)-$\beta$ diagram under the assumption of orientation scenario A. The different colors represent LyC escape fractions $f_\mathrm{esc}=0.9$ (blue), 0.7 (green), 0.5 (orange) and 0.0 (red). {\bf a)} Galaxies obeying the LMC attenuation law {\bf b)} SMC attenuation law {\bf c)} Calzetti attenuation law with $E(B-V)_\mathrm{stars}=0.44E(B-V)_\mathrm{neb}$ {\bf d)} Calzetti $E(B-V)_\mathrm{stars}=E(B-V)_\mathrm{neb}$ \label{EWbetadust_caseA}}
\end{figure*}

Since both the H$\beta$ and H$\gamma$ emission lines are detectable with JWST/NIRSpec at $z=6$--9, and since the intrinsic line ratio is well-constrained from recombination theory, \citet{Zackrisson13} suggested that these lines could be used to correct NIRSpec spectra for dust reddening at these redshifts. However, we find that the modest optical dust attenuation combined with the relatively weak H$\gamma$ line (rest-frame equivalent width $\lesssim$ 10 \AA{} in the case of $f_\mathrm{esc}=0$ and even weaker for higher $f_\mathrm{esc}$) predicted for our simulated galaxies will in most cases make it practically impossible to apply any sensible dust corrections this way. Machine learning methods that exploit the combined marginal detections of several weak spectral features (e.g. H$\gamma$, H$\delta$, H$\epsilon$...) may still be able to provide some handle on the dust reddening, although with substantial uncertainties \citep{Jensen et al.}.  

Figure~\ref{EWbetadust_caseA} illustrates the effect of dust on the  EW(H$\beta$)-$\beta$ diagram for galaxies where the viewing angle corresponds to scenario A (i.e. $f_\mathrm{esc}$-independent attenuation). The different panels show the behavior of S14 galaxies subject to the LMC, SMC, Calzetti $E(B-V)_\mathrm{stars}=E(B-V)_\mathrm{neb}$ and Calzetti $E(B-V)_\mathrm{stars}=0.44E(B-V)_\mathrm{neb}$ attenuation laws. The behavior of galaxies from the other simulation suites subject to the Finlator dust recipe is very similar. The effects of the LMC, SMC and Calzetti $E(B-V)_\mathrm{stars}=E(B-V)_\mathrm{neb}$ attenuation laws (Figure~\ref{EWbetadust_caseA}abd) is to significantly reduce the usefulness of the $\beta$ slope as a diagnostic of LyC leakage, since $\beta$ becomes dominated by dust effects rather than the recent star formation history. Because of this, the $\beta$ distribution becomes very similar for all $f_\mathrm{esc}$ considered, and EW(H$\beta$) may by itself then be used to gauge $f_\mathrm{esc}$. The situation is, however, quite different for the Calzetti $E(B-V)_\mathrm{stars}=0.44E(B-V)_\mathrm{neb}$ case (Figure~\ref{EWbetadust_caseA}c), since dust attenuation in this case not only affects $\beta$, but simultaneously also decreases EW(H$\beta$) by diminishing the relative impact of nebular emission in the spectrum. Galaxies therefore form diagonal distributions in the EW(H$\beta$)-$\beta$ diagram, qualitatively similar to the dust-free case depicted in Figure~\ref{no_dust} but extending to redder $\beta$. 

Inferring $f_\mathrm{esc}$ from the position of a single galaxy in the EW(H$\beta$)-$\beta$ diagram would lead to significant degeneracies without any knowledge on the effective attenuation law. As an example, consider the position of a galaxy at $\beta\approx -2$ and $\log_{10}\ EW(\mathrm{H}\beta)\approx 1.5$. In the LMC, SMC and Calzetti $E(B-V)_\mathrm{stars}=E(B-V)_\mathrm{neb}$ cases, the LyC leakage would be inferred to be $f_\mathrm{esc}\approx 0.5$--0.7, whereas in the $E(B-V)_\mathrm{stars}=0.44E(B-V)_\mathrm{neb}$ case, one would infer $f_\mathrm{esc}\approx 0$--0.5. Information on the effective attenuation law from observations at slightly lower redshifts may help guide the analysis, and so may the distribution of objects at $z>6$ in the EW(H$\beta$)-$\beta$ diagram itself (similarly to the situation discussion in section~\ref{stellar_evolution}), but a case in which the effective attenuation law changes from galaxy to galaxy at $z>6$ would inevitably lead to a very problematic situation. 

\begin{figure}[t]
\centering
\plotone{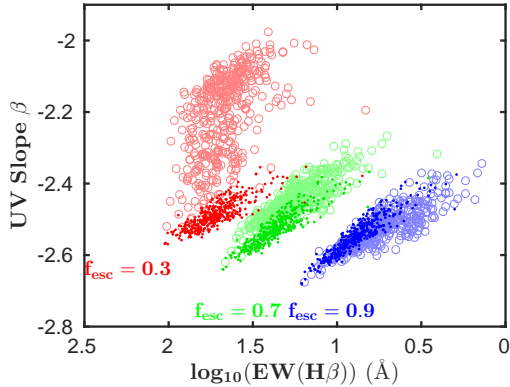}
\caption{Effects of dust attenuation on $z=7$ galaxies from the \citet{Shimizu14} simulations in the EW(H$\beta$)-$\beta$ diagram under the assumption of orientation scenario B ($f_\mathrm{esc}$-dependent dust attenuation) and the LMC attenuation law. The different colors represent LyC escape fractions $f_\mathrm{esc}=0.3$ (red), 0.7 (green) and 0.9 (blue). Open circles represent scenario B and darker dots the completely dust-free case. While the predicted $\beta$ distribution in scenario B approaches that of the dust free-case (identical for both scenarios A and B) at high $f_\mathrm{esc}$, there are lingering effects on the $EW(H\beta)$ distribution because of the contribution to the H$\beta$ continuum from direct, unattenuated starlight, with lower $EW(H\beta)$ as a result.
\label{EWbetadust_caseB}}
\end{figure}

Figure~\ref{EWbetadust_caseB} illustrates the effect of dust on the EW(H$\beta$)-$\beta$ diagram in the alternative scenario B (where unattenuated starlight escapes along the line of sight, leading to $f_\mathrm{esc}$-dependent dust attenuation) for an LMC attenuation law. At $f_\mathrm{esc}=0$ (not shown), scenario B gives predictions identical to those of scenario A in Figure~\ref{EWbetadust_caseA}, but at higher $f_\mathrm{esc}$, scenario B produces SEDs with $\beta$ slopes that are more similar to the completely dust-free case in Figure~\ref{no_dust}. Hence, the $\beta$ slope can in this scenario appear very blue even though there is substantial dust attenuation in the galaxy. This happens because the unattenuated fraction $f_\mathrm{esc}$ of the starlight (which give very blue $\beta$) comes to dominate the rest-frame UV continuum and hence the overall $\beta$ slope. At $f_\mathrm{esc}>0.5$, the predicted distribution of $\beta$ slopes is only marginally different from the dust-free case (here shown as symbols in a darker shade). The EW(H$\beta$) distribution at these very high $f_\mathrm{esc}$ is, on the other hand, expected to be shifted to EW(H$\beta$) slightly lower than in the dust-free case due to a boost in the overall H$\beta$ continuum from the direct, unattenuated starlight. While it would in many cases be impossible to observationally determine which of the scenarios A and B is the most relevant, this is not necessarily a problem for indentifying galaxies with very high levels of LyC leakage ($f_\mathrm{esc}>0.5$), since these would exhibit very low EW(H$\beta$ in both cases. There is, however, one special case where the observables we consider would favour one geometry over the other: Since scenario B galaxies with very high $f_\mathrm{esc}$ always exhibit blue $\beta$ slopes ($\beta<-2.2$), the detection of a galaxy with very low EW(H$\beta$) and a red $\beta$ ($\beta>-2.2$) would suggest that the viewing angle is more similar to scenario A. 

\section{Discussion}
\label{discussion}
In previous sections, we have demonstrated that indirect spectral signatures of LyC leakage in the rest-frame ultraviolet and optical should be readily detectable with JWST/NIRSpec at $z=7$--9, and that individual galaxies with very high LyC escape fractions ($f_\mathrm{esc}\geq 0.5$) should be identifiable at these redshifts based on their position in the $\beta$ vs EW(H$\beta$) diagram. In a companion paper \citep{Jensen et al.}, we also show that the full range of spectral information available from JWST/NIRSpec, low-resolution spectra should allow for constraints with a mean statistical error of $\Delta f_\mathrm{esc}\approx 0.1$. This method of using continuum and emission-line data to constrain LyC leakage from reionization-epoch galaxies has the obvious advantage over competing methods based on absorption lines \citep{Jones et al.,Reddy16,Leethochawalit16} of not requiring high spectral resolution or as high signal-to-noise ratios, which means that it can be applied to much larger samples of $z>6$ galaxies. A recent attempt to test the metal absorption line method using direct LyC measurements at $z\approx 2.4$ also failed to verify the predicted LyC flux \citep{Vasei et al.}, indicating potential problems with this approach.  One can, however, envision similar problems with our proposed method as well. In the following, we will discuss a number of potential pitfalls that require further consideration.

\subsection{Lyman continuum attenuation by dust}
\label{LyC_extinction}

As pointed out by \citet{Zackrisson13}, the method of using the relative strength of emission lines to assess $f_\mathrm{esc}$ would overestimate $f_\mathrm{esc}$ in cases where some non-negligible fraction of the LyC photons that do not escape is directly absorbed by dust rather than gas. For instance, the models presented by \citet{Cai14} suggest that dust attenuation may reduce the fraction of ionizing photons captured by gas by a factor of $\approx 2$ even at $z\approx 6$--9. In terms of the classical geometries often used in photoionization modelling, LyC attenuation by dust occurs in density-bounded nebulae with an outer dust shell, and in both ioniziation-bounded and density-bounded nebulae when photoionized gas and dust are mixed (see Figure 8 in \citealt{Zackrisson13}). LyC attenuation by dust can also happen if dust is sprinkled within the escape channels of an ioinization-bounded nebula with holes, although the simulations of \citet{Gnedin08} indicate that when LyC escapes, it tends to do so through dust-free channels.

Observations of the rest-frame far-IR continuum from dust emission can constrain the level of LyC extinction \citep{Inoue 01,Inoue et al. 00,Inoue et al. 01,Hirashita et al.} and the study by \citet{Borthakur et al.} suggests that this mechanism may have a substantial impact on $f_\mathrm{esc}$ for LyC-leaking galaxies at low redshift. However,  at $z>6$, the dust continuum gets redshifted into the sub-mm regime and is hence outside the wavelength range of JWST. In the case of the very brightest reionization-epoch galaxies, the dust continuum can be probed by ALMA \citep[e.g.][]{Watson et al.}, but the sensitivity limits of ALMA makes this impossible for the vast majority of $z>6$ galaxies within reach of JWST/NIRSpec. ALMA observations of the redshifted dust continuum is far less challenging at $z\approx 3$  \citep{Scoville et al.}, where the opacity of the IGM still allows direct detection of escaping LyC. Hence, it should at least be possible to assess how pronounced LyC extinction is in bright, high-$f_\mathrm{esc}$ galaxies $\approx 1$ Gyr after cosmic reionization has completed. 

\subsection{Heavily obscured star formation}
\label{high_obscuration}
Due to the relatively limited wavelength range within reach of JWST/NIRSpec for galaxies at $z=6$--9 (rest-frame wavelengths 1216 to $\approx 5000$--7100 \AA{}, depending on the exact redshift), it is possible to come up with scenarios that are bound to lead the JWST analysis of such galaxies astray. Objects with very high dust obscuration would represent a particularly troublesome case.

Consider the case where supernovae punch small holes through an extremely dusty and opaque ISM, so that some small fraction $f_\mathrm{esc}$ of starlight is escaping without any attenuation by gas or dust whatsoever, whereas the rest is subject to very high dust attenuation. This resembles our viewing angle scenario B, but with far more pronounced dust effects outside the holes than considered in our simulations. The unobscured part could then completely dominate the rest-frame ultraviolet/optical light. If the unobscured region is young, the observed spectrum would exhibit a very blue UV slope and extremely weak emission lines -- the tell-tale signatures of a very high overall $f_\mathrm{esc}$ -- despite the fact that the global $f_\mathrm{esc}$ is very low. 

In fact, \citet{Kimm13} have proposed a model of this type to explain the observed properties of $z=7$ galaxies, adopting $f_\mathrm{esc}=0.1$ and an optical attenuation of $A_V=1.8$ magnitudes over the attenuated regions. As in the case of direct LyC absorption by dust (Section~\ref{LyC_extinction}), the most direct way to get an observational handle on this scenario would likely be auxiliary observations of the very brightest JWST/NIRSpec targets with ALMA \citep{Cen14}. One may also be able to address this using ensemble statistics. If a significant fraction of star formation is generally completely blocked by dust, then the stellar mass is growing much faster than what is inferred from the rest-frame ultraviolet. By comparing the simultaneous growth of the rest-frame ultraviolet and optical luminosity functions in time, it may then be possible to assess whether this is a dominant mode.

A similar complication could arise for galaxies that have two (or more) blended components with vastly different dust and LyC escape properties. If a high-mass component with $f_\mathrm{esc}=0$ experiences too high attenuation to contribute substantially to the observed JWST/NIRSpec SED, a low-mass, low-obscuration $f_\mathrm{esc}>0$ component could come to dominate the observed SED. In this case, any $f_\mathrm{esc}$ estimate based on our methods would refer solely to the latter component. In a scenario of this kind, one would also severely underestimate the overall stellar mass of the system. 

We stress, however, that very few objects subject to very high obscuration have so far been identified at $z>6$ -- with LFLS3 at $z\approx 6.3$ with $M_\mathrm{stars}\approx 5\times 10^{10}\ M_\odot$ and $A_{UV}\approx 5$ mag \citep{Cooray14} being the only compelling case so far. On the contrary, both observations and simulations suggest that that the average obscuration is $A_{UV}\lesssim 0.5$ mag in all but the most luminous galaxies at $z>6$ \citep[e.g.][]{Wilkins13,Bouwens16}. 

\subsection{Density-bounded nebulae}
\label{density-bounded}
\citet{Zackrisson13} discussed two separate mechanisms for LyC leakage -- an ionization-bounded nebula with holes and a density-bounded nebula. In this paper, we have only considered the former one, since \citet{Zackrisson13} argue that the two should give rise to very similar signatures in the EW(H$\beta$)--$\beta$ diagram. However, the relative strength of emission lines like [OIII]5007 and [OII]3727 (which can both be readily observed with JWST/NIRSpec up to $z\approx 9$) differ in the case of density-bounded nebulae, which means that  JWST/NIRSpec observations should not only be able to constrain $f_\mathrm{esc}$, but also provide the handle on the dominating mode of LyC escape mechanism from galaxies in the reionization epoch. Observations of a galaxy with extreme LyC leakage at $z\approx 3.2$ presented by \citet{Vanzella16} and \citet{deBarros16} already support that the leakage is happening through a density-bounded nebula, and the properties of the interstellar medium of the $z\approx 7$ galaxy studied in [OII]88$\mu$m and [CII]158$\mu$m by \citet{Inoue16} also appear favorable for LyC leakage through this mechanism. For this reason, it would be valuable to extend the present work to include density-bounded nebulae. While this extension is straightforward, it also involves more free parameters, since the oxygen lines are also sensitive to the ionization parameter and the relative oxygen abundance. We do, however, intend to revisit such models in a future paper.

\subsection{Anisotropic LyC leakage}
Simulations suggest that LyC leakage may be highly anisotropic \citep{Paardekooper15,Cen & Kimm}, so that even though the global $f_\mathrm{esc}$ may be significant, the parts of a galaxy that is facing the observer may exhibit very low LyC escape. If so, this would complicate the use of methods aiming to constrain $f_\mathrm{esc}$ using absorption lines, since these are measuring the covering fraction of foreground gas. One could for instance imagine a galaxy in which absorption lines suggest complete coverage and hence zero LyC leakage, whereas extremely high LyC leakage is taking place in the opposite direction. The same problem also plagues direct detections of LyC leakage from individual objects at $z\lesssim 4$, since these are biased by the LyC fraction that happens to be escaping in the direction of the observer.

As discussed in Sections~\ref{SED_modelling} and ~\ref{dust}, our method would be able to detect the signatures of LyC leakage even in the case of highly anisotropic escape, albeit with caveats. In the case of minor dust attenuation, anisotropic LyC leakage would not be a problem, since nebular emission from all sides of the galaxy should be able to reach the observer. Hence, our method should provide a better handle of the {\it global} escape fraction of galaxies. However, scenarios involving high obscuration coupled to anisotropic leakage (as discussed in Section~\ref{high_obscuration}) would introduce a directional bias on the$f_\mathrm{esc}$ inferred by our method as well.

\subsection{Stochastic IMF sampling}
The simulated SEDs presented in this paper are based on assumption that the IMF is fully sampled, which is a safe approximation for high-mass stellar populations. For low-mass systems, however, the incomplete sampling of the IMF may lead to stochastic variations between the ionizing and non-ionizing continuum fluxes, which would complicate attempts to estimate $f_\mathrm{esc}$  based on the equivalent width of emission lines from individual objects. IMF sampling effects are negligible for objects with star formation rates SFR$\gtrsim 1\ M_\odot$ yr$^{-1}$ \citep{Forero-Romero13,daSilva14}, which for our simulated galaxies at $z=7$ correspond to $m_\mathrm{AB}\leq$ 28.5--29.0 at rest-frame wavelengths around 1500 \AA. In the absence of gravitational lensing, JWST/NIRSpec is not likely to be able to push beyond this limit \citep{Jensen et al.}. However, a strongly lensed galaxy that appears at this limit could have SFR$\ll 1\ M_\odot$ yr$^{-1}$, at which point stochastic IMF sampling effects would start to matter. At this point, the method proposed for estimating $f_\mathrm{esc}$ would become unreliable for individual objects. The typical $f_\mathrm{esc}$ in a population of galaxies could in principle still be inferred, but this requires that one takes the predicted distribution of observed properties as a function of $f_\mathrm{esc}$ into account, and corrects for the fact that the galaxies with the most massive stars are more easily detectable. 

\subsection{LyC absorption in low-density circumgalactic gas}
The method for constraining $f_\mathrm{esc}$ in this paper and in \citet{Zackrisson13} is based on notion that the LyC absorbed by gas within galaxies should be reflected in the strength of rest-frame UV/optical emission lines relative to the continuum at these wavelengths (which stems both from direct starlight and free-bound/free-free processes in the interstellar medium). However, in simulations, the radial deliminator at which a LyC photon is considered to have reached the IGM is traditionally placed at the dark halo virial radius, which means that the LyC photon not only needs to escape from the dense interstellar medium within galaxies, but also needs to make it past neutral gas in the circumgalactic medium, which may have much lower density. When the density drops below $n(H)\approx 0.1$ cm$^{-3}$, the hydrogen recombination time scale becomes $\gtrsim 1$ Myr, which means that that the recombination flux per unit time will be much lower than in the denser interstellar medium, and that changes in the stellar LyC output will be reflected in the recombination emission lines with a substantial time delay. As an example of the potential problem that this could introduce, consider an extreme scenario in which a sudden surge of star formation boosts the LyC flux of a stellar population and produces a blue rest-frame stellar ultraviolet continuum, yet all LyC photons are absorbed in low-density, circumgalactic gas. This could result in a situation in which $f_\mathrm{esc}\approx 0$, yet the long recombination timescale ensures extremely weak emission lines, suggesting $f_\mathrm{esc}\approx 1$. Simulations of the covering fraction and density of neutral gas in the circumgalactic medium of galaxies at $z>6$ would be very useful to assess how serious this problem is likely to be. Of course, since the problem is fundamentally due to a mismatch between what is measured in observations and simulations, the issue can be completely evaded by simply redefining the threshold radius at which LyC escape is assumed to have occurred in simulations to provide a better match between the two. 

\section{Summary}
\label{summary}

Our results can be summarized as follows:
\begin{itemize}
\item As part of the LYCAN project, we have generated a large grid of synthetic SED of $z=$7--9 galaxies drawn spectra from cosmological simulations, in a wavelength range relevant for JWST/NIRSpec observations at these redshifts. These models, which are intended for the planning of JWST observations and for training data analysis tools, include a number of different options concerning LyC leakage, stellar evolution, dust attenuation and observational noise, and are publicly available. 

\item We demonstrate, that the method proposed by \citep{Zackrisson13} constrain the LyC leakage of reionization-epoch galaxies using indirect spectral signatures (emission lines and continuum) in JWST/NIRSpec spectra seems to hold even in the light of variations in star formation history and metallicity at the level seen in the simulations. Even very simple combinations of diagnostics like the slope of the UV continuum and the strength of the H$\beta$ emission line allow the identification of extreme ($f_\mathrm{esc}>0.5$) LyC-leaking galaxies (section \ref{diagnostic_nodust}). Broadly speaking, such objects are expected to exhibit rest-frame EW(H$\beta$)$\leq 30$ \AA.

\item Uncertainties in dust attenuation, the effective attenuation law and in stellar evolution (e.g. the role of binary evolution and stellar rotation) at $z>6$ convert into systematic uncertainties in the $f_\mathrm{esc}$ inferred from individual galaxies, but upcoming observations at both low-to-intermediate redshifts and at $z>6$ should allow such model uncertainties to be severely constrained (sections~\ref{diagnostic_nodust}, ~\ref{dust} and \ref{discussion}).

\item Potential complications that warrant further studies include direct LyC absorption by dust, density-bounded nebulae, and the absorption of LyC gas in low-density circumgalactic gas with very long recombination timescales (section~\ref{discussion}). 
\end{itemize}

\acknowledgments
E.Z acknowledges funding from the Swedish Research Council (project 2011-5349). JPP acknowledges support from the European Research Council under the European Communitys Seventh Framework Programme (FP7/2007-2013) via the ERC Advanced Grant ”STARLIGHT: Formation of the First Stars” (project number 339177)

\end{document}